\documentclass[twocolumn,prd,aps,nofootinbib,showpacs,superscriptaddress]{revtex4}
\usepackage{multirow}
\usepackage{graphicx}
\usepackage[dvips]{color}
\usepackage{amssymb,amsmath}

\begin{document}

\title{Lattice QCD study on $K^\ast(892)$ meson decay width}

\author{Ziwen Fu}
\email{fuziwen@scu.edu.cn}

\affiliation{
Key Laboratory of Radiation Physics and Technology {\rm (Sichuan University)},
Ministry of Education; \\
Institute of Nuclear Science and Technology;
College of Physical Science and Technology, Sichuan University,
Chengdu 610064, P. R. China.
}

\author{Kan Fu}
\email{gkanfu@gmail.com}
\affiliation{
School of Environment, Tsinghua University, Peking 100084, P. R. China.
}
\begin{abstract}
We deliver an exploratory lattice QCD examination of
the $K^\ast(892)$ meson decay width with the help of the $p$-wave scattering phase $\delta_1$
of pion-kaon ($\pi K$) system in the isospin $I=1/2$ channel,
which is extracted by the modified Rummukainen-Gottlieb formula
for two-particle system with arbitrary masses,
and it clearly reveals the entity of a resonance at a mass
around $K^\ast(892)$ meson mass.
The effective range formula is applied to describe
the energy dependence of scattering phase
and we obtain the effective $K^\ast \to \pi K$ coupling
constant as $g_{K^\ast \pi K} = 6.38(78)$, and subsequently achieve the decay width
as $\Gamma_{K^\ast}=64.9 \pm 8.0$~MeV, which is in reasonable accordance
with the experiment.
Our lattice investigations are conducted
on a $20^3\times48$ MILC full QCD gauge configuration
at $(m_\pi + m_K) / m_{K^\ast} \approx 0.739$
and the lattice spacing $a \approx 0.15$ fm.
\end{abstract}


\date{\today}
\maketitle

\section{Introduction}
\label{Sec:Introduction}
It is well-known that the vector $K^\ast(892)$ meson is a resonance.
In 2012, the Particle Data Group (PDG)
listed the  $K^\ast(892)$ meson $I(J^P)=\frac{1}{2}(1^-)$,
with a  mass $891.66\pm0.26$~MeV and a narrow width $50.8\pm0.9$~MeV~\cite{Beringer:1900zz}.
Some recent experimental analyses~\cite{Boito:2010me,Mitchell:2009aa,
Aubert:2007ur, Ablikim:2005ni,Bai:2003fv}
have precisely measured its resonance parameters.
Moreover, a few theoretical efforts have been taken to
calculate its hadronic coupling constant~\cite{Gottlieb:1985rc,Gottlieb:1983rh,Vasanti:1976zz}.
Since the $K^\ast(892)$ meson is a low-lying vector meson with strangeness,
a study of its decay width is definitely a straightforward investigation on
the three-flavor structure of the low-energy hadronic interactions,
hence, it is very helpful for us to acquaint 
with the dynamical traits of the hadronic interactions with QCD.

At present, lattice QCD is the most feasible approach from first principles
to extract the resonance parameters of $K^\ast(892)$ meson nonperturbatively.
The principal decay channel
(with a branching rate of $99.9\%$) of $K^\ast(892)$ meson
is to one pion plus one kaon in the $p$-wave~\cite{Beringer:1900zz},
which can then be precisely dealt with on the lattice,
and there is a pioneering lattice QCD study on
its coupling constant $g_{K^\ast \pi K}$ through evaluating appropriate
three-point correlation function~\cite{Loft:1988sy}.
Among unstable hadrons, the vector $\rho$ meson is ideal
(see reasons in~\cite{Feng:2010es})
for lattice QCD investigations of a resonance,
and it is extensively studied~\cite{Loft:1988sy,Altmeyer:1995qx,
McNeile:2002fh,Aoki:2007rd,Ishizuka:2011ic,Aoki:2011yj,
Lang:2011mn,Feng:2010es,Frison:2010ws},
nevertheless, so far, lattice QCD research on
the resonance parameters of $K^\ast(892)$ meson
directly from $p$-wave scattering phase of $\pi K$ system in the $I=1/2$ channel
has not been reported yet, mainly because the rectangular diagram
is exceptionally hard to  rigorously calculate,
the statistical error of the numerically computed
$K^\ast$ mass is not too small, and most of all,
the proper finite size formula to describe $\pi K$ system enclosed in a cubic box
in the moving frame is not completely established yet until recently.

Motivated by the recent extensions and developments of the Rummukainen-Gottlieb
formula~\cite{Rummukainen:1995vs} to a generic two-particle system
with arbitrary masses in the moving frame~\cite{Davoudi:2011md,Fu:2011xz,
Leskovec:2012gb,Gockeler:2012yj,Hansen:2012tf,Bernard:2012bi,Doring:2012eu}
and J.~Nebreda and J.~Pelaez's brilliant expositions
on $K^\ast(892)$ resonance~\cite{Nebreda:2010wv},
and also encouraged by our previous work on
the accurate determination of $K^\ast$ mass~\cite{Fu:2012zk},
the exploratory calculations of the scalar meson decay widths~\cite{Fu:2011xw,Fu:2011xz,Fu:2012gf},
and the accurate computation of the $s$-wave $\pi K$ scattering length
in the $I=1/2$ channel~\cite{Fu:2011wc},
here we will step out further to probe its decay width by way of lattice QCD.

In the present work, we discuss all the possible computation scheme
for calculating the $\pi K$ scattering phase
with total zero momentum in the center-of-mass (CM) frame,
and  total non-zero momentum in the moving frame (MF), respectively,
and obtain the $K^\ast(892)$ decay width by calculating  $p$-wave scattering phase 
of $\pi K$ system in the $I=1/2$ channel in the moving frame.
The calculations are launched on a MILC full QCD gauge configuration
with the $2+1$ flavors of the Asqtad improved staggered quarks~\cite{Bernard:2010fr,Bazavov:2009bb}.
The meson masses quoted from our previous work~\cite{Fu:2012zk}
yielded $(m_\pi+m_K)/m_{K^\ast} \approx 0.739$,
and the lattice ensemble parameters determined
by the MILC collaboration gave the lattice extent $L \approx 3.0$~fm
and the lattice space inverse $1/a=1.373$ GeV~\cite{Bernard:2010fr,Bazavov:2009bb}.
The L\"uscher formula~\cite{Lellouch:2000pv,Luscher:1990ux,Luscher:1990ck}
is, as a usual, applied to the case in the center-of-mass frame,
and we use a newly established finite size formula,
which is the generalization of Rummukainen-Gottlieb formula~\cite{Rummukainen:1995vs}
to the generic two-particle system in the moving
frame~\cite{Davoudi:2011md,Fu:2011xz,Leskovec:2012gb,Gockeler:2012yj,
Hansen:2012tf,Bernard:2012bi,Doring:2012eu},
to estimate the $p$-wave $\pi K$ scattering phase in the $I=1/2$ channel.
The simulations conducted at two energies around the $K^\ast$ resonance mass enable us
to extract the decay with of the $K^\ast(892)$ resonance.

This article is organized as follows.
In Sec.~\ref{Sec:Methods}, we elaborate on our calculation method.
Our concrete lattice calculations
are provided in Sec.~\ref{sec:latticeCal}.
We deliver our results in Sec.~\ref{Sec:Results},
and reach our conclusions and outlooks in Sec.~\ref{Sec:Conclusions}.
Numerical calculations of the zeta function
are courteously supplied in the appendix for reference.

\section{Formalism and method of measurement}
\label{Sec:Methods}
\subsection{The relativistic Breit-Wigner formula}
\label{SubSec:Scattering_Phase}
The $K^\ast(892)$ resonance possesses  quantum numbers $I(J^P)=\frac{1}{2}(1^-)$
and principally decays into one pion and one kaon in the $p$-wave
with a branching rate of $99.9\%$~\cite{Beringer:1900zz}.
For an elastic $\pi K$ scattering in the resonance region,
the relativistic Breit-Wigner formula (RBWF)
for the $p$-wave scattering phase $\delta_1$  can be written as~\cite{Beringer:1900zz}
\begin{equation}
\label{eq:BW}
\tan\delta_1 = \frac{\sqrt{s} \, \Gamma_R(s)}{M_R^2-s} ,  \qquad  s = E_{CM}^2 ,
\end{equation}
where $M_R$ is the resonance position,
$\Gamma_R$ is decay width,
$E_{CM}$ is the center-of-mass energy, and
$s$ is the Mandelstam variable.
The $\Gamma_R(s)$ can be expressed by way of the
effective $K^\ast \rightarrow\pi K$  coupling constant
$g_{K^\ast \pi K}$ as~\cite{Nebreda:2010wv},
\begin{eqnarray}
\label{eq:g_kpK_formula}
\Gamma_R(s) &=& \frac{g^2_{K^\ast \pi K}}{6\pi}\frac{p^3}{s} , \\
p           &=& \frac{1}{2\sqrt{s}}
\sqrt{ \left[s - (m_\pi - m_K)^2\right]\left[ s - (m_\pi + m_K)^2\right] } ,
\nonumber
\end{eqnarray}
Checking equations~(\ref{eq:BW}) and (\ref{eq:g_kpK_formula}),
a representation of the $p$-wave scattering phase
as a function of the invariant mass $\sqrt{s}$ is offered
by the effective range formula (ERF),
\begin{equation}
\label{eq:effective_range_formula}
\tan{\delta_1}=\frac{g^2_{K^\ast \pi K}}{6\pi}
\frac{p^3}{ \sqrt{s}(M_R^2- s)}  ,
\end{equation}
which is applicable in the elastic region
and suits the experimental measurements pretty well.
The ERF permits us a fit or seeking for two unknown quantities:
the coupling constant $g_{K^\ast \pi K}$
and the resonance position $M_R$ from the $p$-wave scattering phases.
Then the $K^\ast$ decay width is computed by
\begin{eqnarray}
\label{eq:decay_width}
\Gamma_{K^\ast} &=&
\Gamma_R(s)\Bigg|_{s=M_R^2} =
\frac{g^2_{{K^\ast}\pi K}}{6\pi}\frac{p_{K^\ast}^3}{M_R^2} , \\
p_{K^\ast} &=&
\frac{1}{2M_R}\sqrt{ [M_R^2 - (m_\pi \hspace{-0.05cm}-\hspace{-0.05cm}m_K)^2]
[ M_R^2 - (m_\pi\hspace{-0.05cm}+\hspace{-0.05cm}m_K)^2] }  .
\nonumber
\end{eqnarray}
Equations~(\ref{eq:effective_range_formula})
and (\ref{eq:decay_width}) provide us an approach
to derive the decay width $\Gamma_{K^\ast}$
by studying the dependence of the $p$-wave $\pi K$ scattering
phase $\delta_1$ on the invariant mass $\sqrt{s}$.
We should stress at this point that we will extensively
apply the ERF approximation in the present study
since the RBWF holds perfectly for relatively narrower objects
and the $K^\ast(892)$ resonance has a pretty narrow decay
width $50.8\pm0.9$~MeV~\cite{Beringer:1900zz}.

\subsection{Finite-volume methods}
In this paper, we only focus on the $\pi K$ system with
the isospin representation of $(I,I_z)=(1/2,1/2)$
and deliberate on the $K^\ast(892)$ meson decay
into one pion plus one kaon in the $p$-wave.

\subsubsection{Center of mass frame}
In the center-of-mass frame, the energy eigenvalues of
the non-interacting $\pi K$ system reads
$$
E = \sqrt{m_\pi^2+ p^2} + \sqrt{m_K^2 + p^2}  ,
$$
where $p=|{\mathbf p}|, \; {\mathbf p}=(2\pi/L){\mathbf n}$, and
${\mathbf n}\in \mathbb{Z}^3$.
The energies for the ${\mathbf n} \ne 0$
are typically larger than the $K^\ast$ resonance mass $m_{K^\ast}$.
For example, the lowest energy for the ${\mathbf n} \ne 0$
calculated from the previous determinations of
$m_\pi$, $m_K$ and $m_{K^\ast}$~\cite{Fu:2012zk}
is $E \approx 1.12\times m_{K^\ast}$,
which is self-evidently not qualified to study the $K^\ast(892)$ meson decay.
Hence, we have no choice but to consider the ${\mathbf n} = 0$ case,
and the energy $E = 0.739 \times m_{K^\ast}$,
which is still not a favorite option.

When considering the interaction between pion and kaon,
the energy eigenstates of $\pi K$ system are displaced
by the hadronic interaction from $E$ to $\overline{E}$,
which are calculated by
$$
\overline{E} = \sqrt{m_\pi^2 + k^2} + \sqrt{m_K^2 + k^2} ,
\quad k=\frac{2\pi}{L}q ,
$$
where the dimensionless momentum $q \in \mathbb{R}$.
Solving this equation for the scattering momentum $k$, we have
$$
k = \frac{1}{2\overline{E}}
\sqrt{[\overline{E}^2 -(m_\pi-m_K)^2][ \overline{E}^2 - (m_\pi+m_K)^2]} .
$$

In this article, we are primarily interested in the energy eigenstates
of $\pi K$ system  in the elastic region
$m_\pi+m_K < \overline{E} < 2(m_\pi+m_K)$.
In the center-of-mass frame these energy eigenstates transform as a vector
(to be specific, the irreducible representation $\Gamma = T_1^+$)
under the cubic group $O_h$.
The $p$-wave $\pi K$ scattering phase
$\delta_1$ is linked to the energy $\overline{E}$
by the L\"uscher formula~\cite{Lellouch:2000pv,Luscher:1990ux,Luscher:1990ck},
\begin{equation}
\label{eq:CMF}
\tan\delta_1(k)=\frac{\pi^{3/2}q}{\mathcal{Z}_{00}(1;q^2)} ,
\end{equation}
where the zeta function is formally defined by
\begin{equation}
\label{eq:Zeta00_CM}
\mathcal{Z}_{00}(s;q^2)=\frac{1}{\sqrt{4\pi}}
\sum_{{\mathbf n}\in\mathbb{Z}^3} \frac{1}{\left(|{\mathbf n}|^2-q^2\right)^s}  .
\end{equation}
The $\mathcal{Z}_{00}(s;q^2)$
has a finite value only when ${\rm Re} \, s > 3/2$,
nevertheless it could be analytically continued to $s = 1$.
We usually evaluate $\mathcal{Z}_{00}(s;q^2)$
using the way described in Ref.~\cite{Yamazaki:2004qb}.
I notice that there exists an equivalent L\"uscher formula in Ref.~\cite{Doring:2011vk},
which is the generalization of the L\"uscher quantization condition
to multiple two-body channels.
Moreover it is easy to calculate and more accurate than L\"uscher formula
in the relativistic case.

\subsubsection{Laboratory frame}
To implement the physical kinematics such that
the energy of $\pi K$ system is pretty close
to $K^\ast$  meson mass,
we recourse to the laboratory frame~\cite{Rummukainen:1995vs},
which is usually called the moving frame.
We have presented the detailed discussions of $\pi K$ system
in the moving frame in Ref.~\cite{Fu:2011xw},
here we just review its essential parts.

Considering a moving frame with non-zero total momentum ${\mathbf P}=(2\pi/L){\mathbf d}$,
${\mathbf d}\in\mathbb{Z}^3$,
the energy eigenvalues of the free pion and koan are 
$$
E_{MF} = \sqrt{m_\pi^2+p_1^2} + \sqrt{m_K^2+ p_2^2} ,
$$
where $p_1=|{\mathbf p}_1|$,
      $p_2=|{\mathbf p}_2|$,
and ${\mathbf p}_1$, ${\mathbf p}_2$ define
the three-momenta of $\pi$ and $K$, respectively,
which meet the periodic boundary condition (PBC),
$$
{\mathbf p}_1=\frac{2\pi}{L}{\mathbf n}_1 , \quad
{\mathbf p}_2=\frac{2\pi}{L}{\mathbf n}_2 , \quad
{\mathbf n}_1,{\mathbf n}_2\in \mathbb{Z}^3,
$$
and total momentum ${\mathbf P}$ satisfies
$
{\mathbf P} = {\mathbf p}_1 + {\mathbf p}_2 .
$

In the center of mass frame,
the energy $E_{CM}$  is
$$
E_{CM} =
\sqrt{m_\pi^2 + p^{*2}} + \sqrt{m_K^2+ p^{*2}}  ,
$$
where total center-of-mass momentum disappears, namely,
$
p^*=| {\mathbf p}^*|, \quad
{\mathbf p}^*={\mathbf p}^*_1=-{\mathbf p}^*_2 ,
$
here we denote the center-of-mass momenta
with an asterisk $(\ast)$~\cite{Rummukainen:1995vs}.
We can  readily verify  that the ${\mathbf p}^*$ are
quantized to the values~\cite{Fu:2011xw}
$$
{\mathbf p}^* =\frac{2\pi}{L}{\mathbf r}\,, \qquad
{\mathbf r} \in P_{\mathbf d} ,
$$
where the set $P_{\mathbf d}$  is
\begin{equation}
\label{eq:set_Pd_MF}
P_{\mathbf d} = \left\{ {\mathbf r} \left|  {\mathbf r} = \vec{\gamma}^{-1}
\left[{\mathbf n}+\frac{{\mathbf d}}{2} \cdot
\left(1+\frac{m_K^2\hspace{-0.1cm}-\hspace{-0.1cm}m_\pi^2}{E_{CM}^2}\right)
\right], \right.  {\mathbf n}\in\mathbb{Z}^3 \right\} ,
\end{equation}
where the boost factor, $\gamma=1/\sqrt{1-{\mathbf v}^2}$,
operates in the direction of the velocity ${\mathbf v}$,
which is calculated from  ${\mathbf v}={\mathbf P}/E_{MF}$,
and for the notational compactness we have taken the shorthand notation~\cite{Rummukainen:1995vs},
\begin{equation}
\label{gamma_factor_sh}
\vec{\gamma}{\mathbf p} =
\gamma{\mathbf p}_{\parallel}+{\mathbf p}_{\perp} ,\qquad
\vec{\gamma}^{-1}{\mathbf p} =
\gamma^{-1}{\mathbf p}_{\parallel}+{\mathbf p}_{\perp} ,
\end{equation}
where ${\mathbf p}_{\parallel}$ and ${\mathbf p}_{\perp}$ are
the ingredients of ${\mathbf p}$ parallel and perpendicular to
the center-of-mass velocity ${\mathbf v}$, respectively~\cite{Rummukainen:1995vs}:
$
{\mathbf p}_{\parallel} =
({\mathbf p}\cdot{\mathbf v}) {\mathbf v}/|\mathbf v|^2,
$
$
{\mathbf p}_{\perp} =  {\mathbf p}-{\mathbf p}_{\parallel}
$.

Using the standard Lorentz transformation,
the energy $E_{CM}$ is connected to the $E_{MF}$  through
$E_{CM} = \gamma^{-1}E_{MF}$,
or by
$E_{CM}^2 = E_{MF}^2-{\mathbf P}^2$.

We are particularly interested in one moving frame:
pion at rest, kaon with the momentum
${\mathbf p} = (2\pi / L) {\mathbf e}_3$ (namely, ${\mathbf d} = {\mathbf e}_3$)
and $K^\ast(892)$ meson with the momentum  ${\mathbf P} = {\mathbf p}$.
For our concrete case, we found that its invariant mass
takes $\sqrt{s} = 0.8788 \times m_{K^\ast}$,
which is significantly closer to $K^\ast$ meson mass $m_{K^\ast}$
than that in the center-of-mass frame.
Finally, we find one suitable to study $K^\ast$ decay, 
and here we solely consider this case.

In the presence of the interaction between pion and kaon,
the $\overline{E}_{CM}$ can be calculated by
$$
\overline{E}_{CM} = \sqrt{m_\pi^2 + k^{2}} + \sqrt{m_K^2 + k^{2}} ,
\quad k = \frac{2\pi}{L} q  .
$$
where the dimensional momentum $q \in \mathbb{R}$.
Solving this equation for the scattering momentum $k$, we arrive at
\begin{equation}
\label{eq:MF_k}
k = \frac{1}{2\overline{E}}
\sqrt{[\overline{E}^2_{CM} - (m_\pi-m_K)^2]
      [\overline{E}^2_{CM} - (m_\pi+m_K)^2]} .
\end{equation}
It is convenient to rewrite equation~(\ref{eq:MF_k}) to an elegant form for later use as
\begin{equation}
\label{eq:MF_k_e}
k^2  = \frac{1}{4}
\left( \overline{E}_{CM} + \frac{m_\pi^2 - m_K^2}{\overline{E}_{CM}} \right)^2 - m_\pi^2  ,
\end{equation}
which is used to calculate the scattering momentum $k$
(including its statistical error),
and investigate the lattice discretization effect.

The energy eigenstates of $\pi K$ system for our moving frame
transform under the tetragonal group $C_{4v}$~\cite{Fu:2011xz}.
Only the irreducible representations $A_1$ and $E$ are associated with
the $p$-wave $\pi K$ scattering states in a torus.
We only compute the energies related with the $A_1$ sector in the present study.
The hadronic interaction displaces the energy eigenstate of $\pi K$ system
from $E$ to $\overline{E}$, and the energy  $\overline{E}$ is
linked to the $p$-wave $\pi K$ scattering phase $\delta_1$ with
the help of newly established finite size formula in the moving frame
for the generic two-particle system with arbitrary masses~\cite{Davoudi:2011md,Fu:2011xz,Gockeler:2012yj},
\begin{equation}
\label{eq:Luscher_MF}
\tan\delta_1(k)=
\frac{\gamma\pi^{3/2}q}{\mathcal{Z}_{00}^{\mathbf d}(1;q^2)
+\frac{2}{\sqrt{5}}q^{-2} \mathcal{Z}_{20}^{\mathbf d}(1;q^2)} ,
\end{equation}
where we does not consider the higher scattering phase shifts
$\delta_l (l=2, 3, 4, \cdots )$~\cite{Leskovec:2012gb},
and the modified zeta functions are formally defined by
\begin{eqnarray}
\mathcal{Z}_{00}^{{\mathbf d}} (s; q^2) &= &\sum_{{\mathbf r}\in P_{\rm d}}
\frac{1} { ( |{\mathbf r}|^2 - q^2)^s }, \cr
\mathcal{Z}^{ \mathbf d }_{ 20 } ( s ; q^2 )&=&
\sum_{ {\mathbf r} \in P_{\mathbf d} } \frac{ r^2 Y_{20}(\Omega_r)  }{( r^2 - q^2 )^s }  ,
\label{zetafunction_MF}
\end{eqnarray}
where $\Omega_r$ represents the solid angle parameters $(\theta, \phi)$ of $\mathbf{r}$
in spherical coordinates and the $Y_{lm}$ are the standard spherical
harmonic functions, and the set $P_{\mathbf d}$  is denoted in Eq.~(\ref{eq:set_Pd_MF}).
The scattering momentum $k$ is calculated
from the invariant mass $\sqrt{s}$ through
$\sqrt{s} = \sqrt{ k^2 + m_\pi^2 } + \sqrt{ k^2 + m_K^2 }$.
We have discussed the numerical calculation method of
the $\mathcal{Z}_{00}^{{\mathbf d}} (1; q^2)$
in Appendix A of Ref.~\cite{Fu:2011xw},
and we will give the numerical calculation method of
the $\mathcal{Z}_{20}^{{\mathbf d}} (1; q^2)$ in Appendix~\ref{appe:zeta}
although there are some general calculations
of the zeta function $\mathcal{Z}_{lm}^{\mathbf d}(s; q^2 )$
in Refs.~\cite{Doring:2011vk,Fu:2011xz,Leskovec:2012gb}.

\subsection{Correlation matrix}
\label{SubSec:Correlation_matrix }
To compute two energy eigenvalues,
i.e., $\overline{E}_n$ ($n=1,2$),
we constitute a $2 \times 2$ correlation function matrix:
\begin{equation}
C(t) = \left(
\begin{array}{ll}
\langle 0 | {\cal O}_{\pi K}^\dag(t)  {\cal O}_{\pi K}(0)  | 0 \rangle &
\langle 0 | {\cal O}_{\pi K}^\dag(t)  {\cal O}_{K^\ast}(0) | 0 \rangle
\vspace{0.3cm} \\
\langle 0 | {\cal O}_{K^\ast}^\dag(t) {\cal O}_{\pi K}(0) | 0 \rangle &
\langle 0 | {\cal O}_{K^\ast}^\dag(t) {\cal O}_{K^\ast}(0)| 0 \rangle
\end{array} \right) ,
\label{eq:CorrMat}
\end{equation}
where ${\cal O}_{K^\ast}(t)$ is an interpolating operator
for the vector $K^\ast(892)$ meson with the  specified momentum
${\mathbf p} = (2\pi / L) {\mathbf e}_3$
and the polarization vector parallel to ${\mathbf p}$;
${\cal O}_{\pi K}(t)$ is an interpolating operator for the $\pi K$ system
with the given  momentum  ${\mathbf p} = (2\pi / L) {\mathbf e}_3$.
These interpolating operators employed in the present work
are actually identical to those in our previous studies~{\cite{Fu:2011wc,Fu:2012zk},
nevertheless, to make this article self-contained,
all the fundamental definitions will be provided in the following as well.

\subsubsection{$\pi K$ sector}
Here we take advantage of the original definitions and
notations~\cite{Nagata:2008wk,Sharpe:1992pp,
Kuramashi:1993ka,Fukugita:1994na,Fukugita:1994ve}
to examine the necessary formulae for the lattice QCD calculation of
the $p$-wave scattering phase of $\pi K$ system enclosed in a torus
at the $I=1/2$ channel.
Let us learn the elastic scattering of
a Nambu-Goldstone pion with zero momentum and
a Nambu-Goldstone kaon with the momentum ${\mathbf p}$
in the Kogut-Susskind (KS) staggered fermion formalism.
Using the interpolating operators ${\cal O}_\pi(x_1), {\cal O}_\pi(x_3)$
for pions at points $x_1, x_3$, and ${\cal O}_K(x_2), {\cal O}_K(x_4)$
for kaons at points $x_2, x_4$, respectively,
the $\pi K$ four-point functions are expressed as\cite{Nagata:2008wk}
$$
C_{\pi K}(x_4,x_3,x_2,x_1) =
\bigl< {\cal O}_K(x_4) {\cal O}_{\pi}(x_3) {\cal O}_K^{\dag}(x_2) {\cal O}_{\pi}^{\dag}(x_1)\bigr>  .
$$
where the pion and kaon interpolating field operators are denoted by\cite{Nagata:2008wk}
\begin{eqnarray}
{\cal O}_{\pi^+}({\mathbf{x}},t) &=&
- \overline{d}({\mathbf{x}},t)\gamma_5 u({\mathbf{x}},t)  , \cr
{\cal O}_{\pi^0}({\mathbf{x}},t) &=&
\frac{1}{\sqrt{2}}
[\overline{u}({\mathbf{x}},t)\gamma_5 u({\mathbf{x}},t) -
 \overline{d}({\mathbf{x}},t)\gamma_5 d({\mathbf{x}},t) ]  , \cr
{\cal O}_{K^0}({\mathbf{x}},t)   &=&
\overline{s}({\mathbf{x}},t)\gamma_5 d({\mathbf{x}}, t)  , \cr
{\cal O}_{K^+}({\mathbf{x}},t)   &=&
 \overline{s}({\mathbf{x}},t)\gamma_5 u({\mathbf{x}},t)  .
\nonumber
\end{eqnarray}
After carrying out the summation over the spatial
coordinates $\mathbf{x}_1$, $\mathbf{x}_2$, $\mathbf{x}_3$ and $\mathbf{x}_4$,
we gain $\pi K$ four-point function with the momentum ${\mathbf p}$
as~\cite{Fu:2011xw}
\begin{eqnarray}
\label{EQ:4point_pK_mom}
C_{\pi K}({\mathbf p}; t_4,t_3,t_2,t_1) &=&
\sum_{\mathbf{x}_1} \sum_{\mathbf{x}_2} \sum_{\mathbf{x}_3} \sum_{\mathbf{x}_4}
e^{ i{\mathbf p} \cdot ({\mathbf{x}}_4 -{\mathbf{x}}_2) } \cr
&&\times C_{\pi K}(x_4,x_3,x_2,x_1)  ,
\nonumber
\end{eqnarray}
where $x_1 \equiv ({\mathbf{x}}_1,t_1)$,  $x_2 \equiv ({\mathbf{x}}_2,t_2)$,
$x_3 \equiv ({\mathbf{x}}_3,t_3)$, and $x_4 \equiv ({\mathbf{x}}_4,t_4)$.
To  refrain the color Fierz rearrangement of the quark lines~\cite{Fukugita:1994ve},
we choose $t_1 \ne t_2 \ne t_3 \ne t_4$ 
and  set $t_1 =0, t_2=1, t_3=t$, and $t_4 = t+1$, respectively,
here $t$ represents the time difference.
We build the $\pi K$ interpolating operator in the $I=1/2$ channel as~\cite{Nagata:2008wk}
\begin{eqnarray}
\label{EQ:op_pipi}
{\cal O}_{\pi K}^{I=\frac{1}{2}}({\mathbf{p}}, t) &=& \frac{1}{\sqrt{3}}
\Bigl\{  \sqrt{2}\pi^+(t) K^0({\mathbf{p}}, t+1) \cr
&&-  \pi^{0}(t) K^{+}({\mathbf{p}}, t+1)  \Bigl\}  ,
\end{eqnarray}
where ${\mathbf{p}}$ is total momentum  of $\pi K$ system.
The $\pi K$ operator has the isospin representation with $(I,I_z)=(1/2, 1/2)$.

Considering that $u$ and $d$ quarks have the equal mass,
topologically only three quark line diagrams still contribute to
$\pi K $ scattering amplitudes~\cite{Nagata:2008wk}.
These diagrams are elucidated in Fig.~\ref{fig:diagram},
and we usually label them as direct (D), crossed  (C)
and rectangular (R) diagrams,
respectively~\cite{Kuramashi:1993ka,Fukugita:1994ve}~\footnote{
In Ref.~\cite{Nagata:2008wk},
they are denoted as  $A$, $H$, and $X$, respectively.
}.
The direct and crossed diagrams can be readily computed~\cite{Kuramashi:1993ka,Fukugita:1994ve}
by means of only two fixed wall sources placed at the time slices $t_1$ and $t_2$,
which enables a relatively cheap lattice calculation of the $I=3/2$ $\pi K$  scattering length~\cite{Miao:2004gy,Beane:2006gj}.
Nevertheless, the rectangular diagram (R) needs extra quark propagator
linking the time slices $t_3$ and $t_4$,
which make the strict  evaluation of this diagram extraordinarily expensive.
\vspace{0.8cm}
\begin{figure}[h!]
\includegraphics[width=8.0cm]{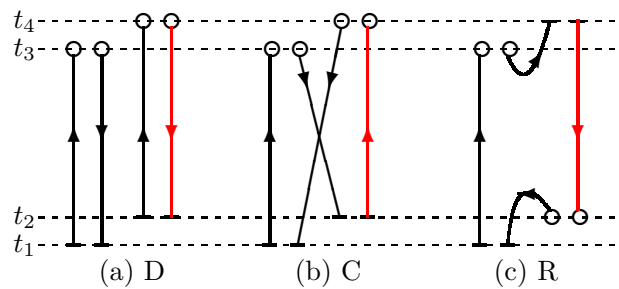}
\caption{ \label{fig:diagram} (color online).
Quark-link diagrams contributing to $\pi K$ four-point functions.
Short black bars stand for the wall sources.
Open circles are sinks for local pion or kaon operators.
The black lines represent for the $u/d$ quark lines,
and red  lines stand for the strange quark lines.
}
\end{figure}

Sasaki et al. handled this puzzle through the technique
with a fixed kaon sink operator to save the computational resources~\cite{Sasaki:2010zz}.
Lang et al. recently solved this problem by the use of Laplacian-Heavyside smeared quarks
within the distillation method~\cite{Lang:2012sv}~\footnote{
It is well-known that the rectangular diagram
(or backtracking contractions, box diagram~\cite{Lang:2012sv})
is most challenging and important for the $I=1/2$ channel.
And obtaining the reliable signal of it is vital to  our final result.
We observe that the signals of the rectangular diagram in Ref.~\cite{Lang:2012sv}
are at reasonable levels.
}.
In our previous works~\cite{Fu:2011xz,Fu:2011xw}, we settled the question
using the moving wall sources without gauge fixing
introduced first by Kuramashi et al.~\cite{Kuramashi:1993ka,Fukugita:1994ve}.
More specifically, we calculated these diagrams by computing each $T$ quark propagators
corresponding to the wall source at all the time slice~\cite{Kuramashi:1993ka,Fukugita:1994ve,Fu:2011xz,Fu:2011xw},
$$
\sum_{n''}D_{n',n''}G_t(n'') = \sum_{\mathbf{x}}
\delta_{n',({\mathbf{x}},t)}, \quad 0 \leq t \leq T-1 ,
$$
where $D$ is the Dirac quark matrix,
and the subscript $t$ in the quark propagator $G$ indicates
the position of the wall source in the temporal direction~\cite{Kuramashi:1993ka,Fukugita:1994ve,Fu:2011xz,Fu:2011xw}.
The associations of the quark propagators $G_t(n)$
exploiting for the $\pi K$ four-point correlation functions
are schematically illustrated in Fig.~\ref{fig:diagram}~\cite{Kuramashi:1993ka,Fukugita:1994ve,Fu:2011xz,Fu:2011xw}.
As we practiced in Ref.~\cite{Fu:2011xw} for the study of $\kappa$ decay width,
for the non-zero momentum $\mathbf{p}$,
we take an up quark source with $1$, and a
strange quark source with the $e^{i{\mathbf p} \cdot {\mathbf{x}} }$
on each lattice site for the pion and kaon creation operator, respectively~\cite{Fu:2011xw}.
By means of the quark propagators $G$,
we can represent $D$, $C$, and $R$ diagrams as~\cite{Fu:2011xw}
\begin{widetext}
\begin{eqnarray}
\label{eq:dcr}
C^D_{\pi K}({\mathbf p};t_4,t_3,t_2,t_1)
&=& \sum_{\mathbf{x}_3} \sum_{\mathbf{x}_4}
e^{ i{\mathbf p} \cdot {\mathbf{x}}_4 }
\langle \mbox{Tr}
[G_{t_1}^{\dag}({\mathbf{x}}_3,t_3)G_{t_1}({\mathbf{x}}_3,t_3)]
\mbox{Tr}
[G_{t_2}^{\dag}({\mathbf{x}}_4,t_4)G_{t_2}({\mathbf{x}}_4,t_4)] \rangle,\cr
C^C_{\pi K}({\mathbf p};t_4,t_3,t_2,t_1)
&=& \sum_{\mathbf{x}_3} \sum_{\mathbf{x}_4}
e^{ i{\mathbf p} \cdot {\mathbf{x}}_4 }
\langle \mbox{Tr}
[G_{t_1}^{\dag}({\mathbf{x}}_3, t_3) G_{t_2}({\mathbf{x}}_3, t_3)
 G_{t_2}^{\dag}({\mathbf{x}}_4, t_4) G_{t_1}({\mathbf{x}}_4, t_4) ] \rangle,\cr
C^R_{\pi K}({\mathbf p}; t_4,t_3,t_2,t_1)
&=& \sum_{\mathbf{x}_2} \sum_{\mathbf{x}_3}
e^{ i{\mathbf p} \cdot {\mathbf{x}}_2 }
\langle \mbox{Tr}
[G_{t_1}^{\dag}({\mathbf{x}}_2, t_2) G_{t_4}({\mathbf{x}}_2, t_2)
 G_{t_4}^{\dag}({\mathbf{x}}_3, t_3) G_{t_1}({\mathbf{x}}_3, t_3) ] \rangle,
\end{eqnarray}
\end{widetext}
where the traces are taken over color,
and the hermiticity natures of the quark propagator $G$ have been applied
to remove the $\gamma^5$ factors~\cite{Kuramashi:1993ka,Fukugita:1994ve,Fu:2011xz,Fu:2011xw}.

As discussed in Refs.~\cite{Kuramashi:1993ka,Fukugita:1994ve},
the rectangular diagram $R$ produce the gauge-variant noise,
and we usually reduce it by conducting the gauge field average without gauge fixing
as we practiced in Refs.~\cite{Fu:2011xz,Fu:2011xw,Fu:2012ng,Fu:2011bz}.
All the three quark line diagrams in Fig.~\ref{fig:diagram} are needed
to compute the $p$-wave $\pi K$ scattering phase in the $I=1/2$ channel.
In the isospin limit, the $\pi K$ four-point function
in the $I=1/2$ channel can be described in terms of only three quark line diagrams~\cite{Nagata:2008wk}, 
\begin{eqnarray}
\label{EQ:phy_I12_32}
\hspace{-2cm}C_{\pi K}({\mathbf p}, t)
&\equiv&
\left\langle {\cal O}_{\pi K}({\mathbf p}, t) |
             {\cal O}_{\pi K}({\mathbf 0}, 0) \right\rangle \cr
\hspace{-2cm}&=&
D + \frac{1}{2} N_f C - \frac{3}{2} N_f R  ,
\end{eqnarray}
where the interpolating field operator ${\cal O}_{\pi K}$ denoted in Eq.~(\ref{EQ:op_pipi})
generates a $\pi K$ state with the total isospin $1/2$
and momentum ${\mathbf p}$,
and $N_f$ is the staggered-flavor factor,
which  is plugged in to address for
the flavor degrees of freedom of the KS staggered fermion~\cite{Sharpe:1992pp}.
We should bear firmly in mind that if we carry out the appropriate root of
the staggered fermion determinant~\footnote{
There are some strong evidences to demonstrate that
the contribution from a single Dirac fermion can be nicely restored
by carrying out the fourth root of the fermion determinant,
see more details in Ref~\cite{degrandANDdetar}.
In the present work, we suppose that the fourth-root procedure
reproduces the proper continuum limit of QCD,
and the lattice results of this work rely on this hypothesis.
Please consult Refs.~\cite{Durr:2004as,Bernard:2006zw,
Bernard:2006vv,Creutz:2007nv,Bernard:2006ee,Creutz:2007yg,Durr:2004ta,
Durr:2006ze,Hasenfratz:2006nw} for the recent investigations about the fourth-root recipe.
}, in the continuum limit,
the same number of the flavors flow around the internal quark loops
as in QCD~\cite{Sharpe:1992pp}.
Therefore, at the level of these quark line diagrams (namely, $D$, $C$, and $R$),
all contributions are exactly as in QCD~\cite{Sharpe:1992pp}.

In practice, we compute the ratios as well~\footnote{
If imposing the Dirichlet boundary condition in the temporal direction,
we can easily extract the energy shift $\delta E$ from
ratios $R^X$~\cite{Kuramashi:1993ka,Fukugita:1994ve}.
On the other hand, we can  readily check that when $t \ll T/2$,
even we set the PBC in the temporal direction,
we still can roughly estimate $\delta E$ from these ratios.
}
\begin{equation}
\label{EQ:ratio}
R^X(t) = \frac{ C_{\pi K}^X({\mathbf p}; 0,1,t,t+1) }
{ C_\pi ({\mathbf 0}; 0,t) C_K({\mathbf p}; 1,t+1) },
\quad  X = D, C, \ {\rm and} \ R  ,
\end{equation}
where $C_\pi ({\mathbf 0}; 0,t)$ and
$C_K ({\mathbf p}; 1,t+1)$ are
the $\pi$ and $K$ correlators with the momentum
${\mathbf 0}$ and ${\mathbf p}$,  respectively.

We should bear in memory that the dedications of non-Nambu-Goldstone
pion and kaon in the intermediate states are exponentially
reduced for the large time owing to their heavier masses
compared with these of Nambu-Goldstone pion and kaon~\cite{Sharpe:1992pp,Kuramashi:1993ka,Fukugita:1994ve,Fu:2011xz,Fu:2011xw}.
Thus, we can ignore this systematic error
due to other $\pi K$  tastes.

\subsubsection{$K^\ast(892)$ sector}
In principle, we can calculate the propagators for two local vector $K^\ast$ meson,
$\gamma_i \otimes \gamma_i$ (VT)  and
$\gamma_0 \gamma_i \otimes \gamma_0\gamma_i$ (PV)\cite{Bernard:2001av,Aubin:2004wf}.
However, here we simply quote the results for local
VT  $K^\ast$ meson since it delivers quite stable results
in the analysis of the mass spectrum.
Moreover, the numerical evaluation of $K^\ast \to \pi K$
three-point function is much eased
if we adopt local VT $K^\ast$ operator.
Thus, we use an interpolation operator
with the isospin $I=1/2$ and $J^{P}=1^{-}$
at the source and sink~\cite{Fu:2012zk}, namely,
$$
{\cal O}(x)  \equiv
\sum_{a} u_a(x) \gamma_i \otimes \gamma_i \bar{s}_a( x ) ,
$$
where $a$ is the color index.
The time slice correlator for the $K^\ast$ meson
in the momentum ${\bf p}$ state is computed by
\begin{eqnarray}
C_{K^\ast}({\mathbf{p} }, t) &=& \sum_{\mathbf{x}} \sum_{ a, b}
e^{i{\mathbf p} \cdot {\mathbf x} }
\langle u_b( {\mathbf{x} }, t) \gamma_i \otimes \gamma_i \bar s_b ({\mathbf{x}}, t)  \cr
&&\times
s_{a}({\mathbf 0}, 0) \gamma_i \otimes \gamma_i \bar u^{a}_{g }({\mathbf 0}, 0)  \rangle ,
\nonumber
\end{eqnarray}
where ${\mathbf 0}, \mathbf{x}$ are
the spatial points of the $K^\ast$ state at source and sink, respectively.

For the staggered quarks, the meson correlators
 have the general single-particle representation,
$$
{\cal C}(t) =
\sum_i A_i e^{-m_i t} + \sum_i A_i^{\prime}(-1)^t e^{-m_i^\prime t}  +(t \rightarrow N_t-t) ,
$$
where the oscillating terms correspond to a meson with the opposite parity.
For $K^\ast$ meson correlator,
we take only one mass with each parity,
and the oscillating parity partner is the $p$-wave meson
with the $J^P=1^+$. The $K_1$ meson is with $J^P=1^+$,
so it is the candidate of the oscillating parity partner
of the vector $K^\ast$ meson.
However, these states with $J^P=1^+$ can just as well be multihadron states~\footnote{
Private communication, C. DeTar (2012).
}.
With staggered fermions, the multihadron possibilities include the various
taste combinations. So we can not identify
its parity partner with the $K_1$, see more discussions in Ref.~\cite{Fu:2012zk}.
Thus, the $K^\ast(892)$ correlator was fit to 
\begin{equation}
\label{eq:kfit}
C_{K^\ast}(t)  = b_{K^\ast}e^{-m_{K^\ast}t} +
b_{K_1}(-1)^t e^{-M_{K_1}t} + (t \rightarrow N_t-t),
\end{equation}
where $b_{K_1}$ and $b_{K^*}$ are two overlap factors.

\subsubsection{ Off-diagonal sector}
A calculation of the generic three-point function
are briefly discussed in  Ref.~\cite{Loft:1988sy}.
To rigorously evaluate it we must compute a spatial volume number of
propagators, namely $N_L^3$ ($16^3$ for our case).
To avoid the apparent intractability of the exactly computing this problem,
S. Golttlieb et al. introduced the ``exponential'' method,
which calculate a two-point function with the presence of a source,
and then differentiates with the source strength to achieve the corresponding
three-point functions~\cite{Gottlieb:1985rc,Gottlieb:1983rh}.
To investigate vector meson decay into pseudoscalars from quenched
lattice QCD~\cite{Loft:1988sy}, Loft and DeGrand adopted ``two-stage''
technique~\cite{Bernard:1985ss,Bernard:1985tm},
which  takes approximately twice as compared with
the calculation of the mass spectra~\cite{Loft:1988sy}.
Later, when studying the resonance parameter of the vector $\rho$ meson~\cite{Loft:1988sy,Altmeyer:1995qx,McNeile:2002fh,Aoki:2007rd,
Ishizuka:2011ic,Aoki:2011yj,Lang:2011mn,Feng:2010es,Frison:2010ws},
people chiefly  employ a stochastic method~\cite{Drummond:1982sk,Dong:1993pk,Foster:1998vw}
or its variants to evaluate three-point correlation function.

Motivated by the precisely evaluate the $\pi\pi$ four-point correlation
functions by Kuramashi et al~\cite{Kuramashi:1993ka,Fukugita:1994ve}
with the moving wall source technique~\cite{Fu:2011wc},
analogously, we have successfully extended
this technique to evaluate three-point correlation function,
and obtained  pretty good signals for the three-point functions
of the $\pi\pi \to \sigma$~\cite{Fu:2012gf}
and $\pi K \to \kappa$~\cite{Fu:2011xw}.
In this work we will continue to use this technique to evaluate
the $\pi K \to K^\ast$ three-point correlation function.

To prevent the tangled color Fierz transformation of
the quark lines~\cite{Fukugita:1994ve},
we should choose $t_1 \ne t_2 \ne t_3$.
In practice, we pick $t_1 =0, t_2=1$, and $t_3=t$
for the $\pi K \to K^\ast$ three-point correlation function,
and opt $t_1 =0, t_2=t$, and $t_3=t+1$
for the $K^\ast \to \pi K$ three-point correlation function.
The quark line diagrams corresponding to the $K^\ast \to \pi K$ and
$\pi K \to K^\ast$  three-point functions are schematically illustrated
in Fig.~\ref{fig:3diagram}(a) and Fig.~\ref{fig:3diagram}(b), respectively.

\vspace{0.3cm}
\begin{figure}[th]
\begin{center}
\includegraphics[width=6.5cm]{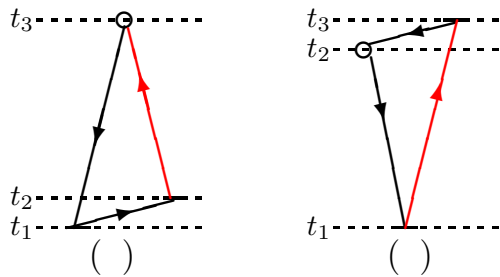}
\end{center}
\vspace{-0.3cm}
\caption{ \label{fig:3diagram}
(color online). Quark-link diagrams contributing to $\pi K \to K^\ast$ and
$K^\ast \to \pi K$ three-point correlation functions.
Short black bars stand for the wall sources.
The black lines represent for the $u/d$ quark lines,
and the red  lines stand for the strange quark lines.
(a) Quark contractions of $\pi K \to K^\ast$,
where open circle is the sink for local $K^\ast$ operator.
(b) Quark contractions of $K^\ast \to \pi K$,
where open circle is the sink for local pion operator.
}
\end{figure}

The $\pi K \to K^\ast$  three-point function can be easily
evaluated with only two fixed wall sources~\cite{Fu:2011xz,Fu:2011xw}.
Nevertheless, the computation of $K^\ast \to \pi K$
three-point function is pretty hard,
since it requires the extra quark propagator
linking time slices $t_2$ and $t_3$~\cite{Fu:2011xz,Fu:2011xw}.
In practice, we employ an up quark source with $1$ on each site
for pion creation operator,
and a strange quark source with $e^{i{\mathbf{p}}\cdot{\mathbf{x}}}$
on each lattice site for kaon creation operator~\cite{Fu:2011xw}.
We write the $K^\ast \to \pi K$ and $\pi K \to K^\ast$  three-point functions
in terms of the quark propagators $G$,
\begin{eqnarray}
\label{eq:dcr3}
\hspace{-0.5cm} C_{\pi K \to K^\ast} ({\mathbf p};t_3,t_2,t_1)
\hspace{-0.15cm}&=&\hspace{-0.15cm}
\sum_{ {\mathbf{x}}_3, {\mathbf{x}}_2}
e^{ i {\mathbf p} \cdot {\mathbf{x}}_3 } \langle  \mbox{Tr}
[G_{t_2}({\mathbf{x}}_3, t_3) \gamma_5 \cr
&& \times G_{t_1}^{\dag}({\mathbf{x}}_3, t_3) \gamma_3
          G_{t_1} ({\mathbf{x}}_2, t_2) ] \gamma_5  \rangle , \cr
\hspace{-0.5cm}  C_{K^\ast \to \pi K}({\mathbf p};t_3,t_2,t_1)
\hspace{-0.15cm}&=&\hspace{-0.15cm}
\sum_{ {\mathbf{x}}_2, {\mathbf{x}}_3}
e^{ i {\mathbf p} \cdot {\mathbf{x}}_2 } \langle  \mbox{Tr}
[G_{t_3}({\mathbf{x}}_2, t_2) \gamma_3 \cr
\hspace{-0.15cm}&&\hspace{-0.15cm}\times
G_{t_1}^{\dag}({\mathbf{x}}_2, t_2) \gamma_5
 G_{t_1} ({\mathbf{x}}_3, t_3) ] \gamma_5 \rangle ,
\end{eqnarray}
where trace is over the color index; the Dirac matrix
are used as an interpolating field for the $i$th meson:
$\gamma_5$ for pseudoscalars and $\gamma_3$ for the vector meson.

\subsection{ Extraction of energies }
\label{SubSec:Extraction of energies}
To map out ``avoided level crossings'' between the $K^\ast$ resonance
and its decay products,
it is important to use the variational method~\cite{Luscher:1990ck} 
to separate the ground state from the first excited state
by calculating a $2 \times 2 $ correlation function matrix
$C(t)$ denoted in~(\ref{eq:CorrMat}).
For this purpose, we construct a ratio of the correlation function matrices as
\begin{equation}
M(t,t_R) = C(t)  \, C^{-1}(t_R)  ,
\label{eq:M_def}
\end{equation}
with some reference time  $t_R$~\cite{Luscher:1990ck}
to extract two lowest energy eigenvalues $\overline{E}_n$ ($n=1,2$),
which can be obtained by a cosh-fit to two eigenvalues
$\lambda_n (t,t_R)$ ($n=1,2$) of the correlation matrix $M(t,t_R)$.
Considering the use of the staggered fermion,
it is easy to verify that $\lambda_n (t,t_R)$ ($n=1,2$) explicitly
has an oscillating term~\cite{Barkai:1985gy,Mihaly:1996ue,Mihaly:Ph.D},
\begin{eqnarray}
\label{Eq:asy}
\lambda_n (t, t_R) &=&
       A_n \cosh\left[-E_n\left(t-\frac{T}{2}\right)\right] \cr
       &&+
(-1)^t B_n \cosh\left[-E_n^{\prime}\left(t-\frac{T}{2}\right)\right] ,
\end{eqnarray}
for a large $t$, which mean $0 \ll t_R < t \ll T/2$ to
suppress both the excited states and wrap-around
contributions~\cite{Fu:2011xw,Gupta:1993rn,Feng:2009ij,Umeda:2007hy}~\footnote{
In Ref.~\cite{Fu:2011xw}, we gave a detailed discussion
about a contamination  from ``wraparound'' effects.
In practice, we will select the fitting time ranges
satisfying $t_{\rm max} \le 16$, and reasonably neglect it.
}.
Without loss of generality, we suppose $\lambda_1(t,t_R) > \lambda_2(t,t_R)$.

\section{Lattice calculation}
\label{sec:latticeCal}
\subsection{Simulation parameters}
We use the MILC full QCD gauge configurations in the presence of
the $N_f=2+1$ flavors of the Asqtad-improved
staggered fermions~\cite{Aubin:2004wf,Bernard:2001av}
and a Symanzik-improved gluon action~\cite{Alford:1995hw}.
We should keep in memory that the MILC gauge configurations are
generated using the staggered formulation of lattice
fermions~\cite{Kaplan:1992bt} with the fourth root of the
fermion determinant~\cite{Bernard:2001av}.

We measured the $\pi K$ four-point correlation functions on the $0.15$ fm
MILC ``medium'' coarse lattice ensemble of $400$ $20^3 \times 48$ gauge configurations
with the bare quark masses $am_{ud} = 0.00484$
and $am_s = 0.0484$ and bare gauge coupling $10/g^2 = 6.566$.
The inverse lattice spacing $a^{-1}=1.373^{+34}_{-14}$ GeV and 
the lattice extent $L\approx3.0$~fm~\cite{Bernard:2010fr,Bazavov:2009bb}.
The mass of the dynamical strange quark is quite close to its physical value,
and the masses of the $u$ and $d$ quarks are
degenerate~\cite{Bernard:2010fr,Bazavov:2009bb}.
The more detailed descriptions of the simulation parameters can be found in Refs.~\cite{Bernard:2010fr,Bazavov:2009bb}.
The PBC is imposed to three spatial directions and temporal direction.

\subsection{Computations}
We employ the standard conjugate gradient method to achieve
the necessary matrix element of the inverse Dirac fermion matrix
to compute the $\pi K$ four-point functions.
We compute the correlators on all the time slices,
and explicitly combine the results from each of the $N_T=48$ time slices.
To be specific, the diagonal correlator $C_{11}(t)$ is measured through
\begin{eqnarray}
 C_{11}(t) &=& \left\langle
\left(\pi K\right)(t)\left(\pi K\right)^\dag(0) \right\rangle \cr
&=&
\frac{1}{T}\sum_{t_s}\left\langle
\left(\pi K\right)(t+t_s)\left(\pi K\right)^\dag(t_s)\right\rangle .
\nonumber
\end{eqnarray}
After averaging the propagator  over all $N_T=48$ possible values,
we found that the statistics are significantly improved.

For each time slice, six Dirac fermion matrix inversions are
needed to compute for the possible $3$ color choices
for the pion source and kaon source, respectively.
So, totally we carry out $288$ matrix inversions
on a single gauge configuration.
This big number of the matrix inversions, conducted on $400$ MILC gauge configurations,
furnishes the gigantic statistics required to
precisely compute the $\pi K$ four-point correlation functions.

For the another diagonal correlator $C_{22}(t)$, $K^\ast(892)$ correlator,
we simply exploit the available propagators measured in our previous study~\cite{Fu:2012zk} to calculate the $K^\ast(892)$ correlator
$$
C_{22}(t)=\frac{1}{T}\sum_{t_s}
\left\langle{K^\ast}^\dag(t+t_s) K^\ast(t_s)\right\rangle ,
$$
where we sum the correlator over
all the time slices  and average it too.

We evaluate the first off-diagonal correlator $C_{21}(t)$:
the $\pi K \to K^\ast$  three-point function, through
\begin{eqnarray}
C_{21}(t) &=& \left\langle K^\ast(t)(\pi K)^\dag(0)\right\rangle \cr
&=&
\frac{1}{T}\sum_{t_s}
\left\langle K^\ast(t+t_s)(\pi K)^\dag(t_s)\right\rangle ,
\nonumber
\end{eqnarray}
where the summation is  over all the time slice.
Through the relation $C_{12}(t)=C_{21}^\ast(t)$,
we can gratuitously gain the second off-diagonal correlator $C_{12}(t)$:
the $K^\ast \to \pi K$  three-point function.

In the present study, we evaluate two-point correlation functions
for pion and kaon as well,
\begin{eqnarray}
G_\pi({\mathbf 0};t) &=&  \frac{1}{T}  \sum_{t_s}
\langle 0|\pi^\dag ({\mathbf 0},t+t_s) \pi({\mathbf 0},t_s) |0\rangle  , \cr
G_K({\mathbf p};t)   &=&  \frac{1}{T}  \sum_{t_s}
\langle 0|  K^\dag ({\mathbf p},t+t_s)   K({\mathbf p},t_s) |0\rangle  ,
\label{eq:Gpi}
\end{eqnarray}
where the summation is  over all the time slice,
and  the $G_\pi ({\mathbf 0};t)$, $G_K ({\mathbf p};t)$
are the two-point correlation functions
for pion meson with zero momentum,
and kaon meson with the momentum ${\mathbf p}$, respectively.

\section{Simulation results }
\label{Sec:Results}

In our previous work~\cite{Fu:2012zk},
we have measured the  point-to-point pion and kaon  correlators with high accuracy.
Exploiting these correlators, we can precisely derive
the pion mass ($m_\pi$) and kaon mass ($m_K$),
which are in fair agreement with the previous MILC
determinations in Ref.~\cite{Bazavov:2009bb}.
In Table~\ref{table:Disp} we list the pion mass $m_\pi$,
the mass $m_K$ and energy $E_K$ of kaon meson
with the momentum ${\mathbf p}=(2\pi/L){\mathbf e}_3$,
which are extracted through a single exponential fit ansatz to
$G_\pi (t;{\mathbf 0})$ and $G_K(t;{\mathbf p})$ in Eq.~(\ref{eq:Gpi}).
We show the mass $m_{K^\ast}$ and energy $E_{K^\ast}$
of the vector $K^\ast$ meson with the momentum ${\mathbf p}=(2\pi/L){\mathbf e}_3$
as well, which are extracted from the $K^\ast(892)$  correlator.
\begin{table}[h]
\caption{\label{table:Disp}
Masses $m$ of the pion, kaon and $K^\ast(892)$ mesons,
and energies $E$ of kaon  and $K^\ast(892)$ mesons
with the momentum ${\mathbf p} = (2\pi/L) {\mathbf e}_3$,
extracted from the corresponding point-to-point correlation functions.
}
\begin{ruledtabular}
\begin{tabular}{ c l l c }
     & $\pi$           & K  &  $K^\ast(892)$  \\
\hline
$am$  &  $ 0.17503(17) $  & $0.39913(27) $ & $0.7757(70)$  \\
$aE$  &                   & $0.50465(48) $ & $0.8278(82)$  \\
\end{tabular}
\end{ruledtabular}

\end{table}

We must stress at this point that, in the present work,
we just use this calculated $K^\ast(892)$ mass $m_{K^\ast}$
to indicate the position of free $K^\ast(892)$  mass,
which are marked by the fancy cyan plus point in Fig.~\ref{fig:sin2Del},
and we visualize this value to compare with the resonance mass $M_R$.

\subsection{Diagrams D, C, and R}
The individual ratios $R^X$ ($X=D, C$ and $R$),
which correspond to the diagrams in Fig.~\ref{fig:diagram},
are illustrated in Fig.~\ref{fig:ratio} as the functions of time separation $t$.
The values of the direct amplitude ratio $R^D$ are pretty close to unity,
implying a quite slight interaction in this channel.
On the other hand, the crossed amplitude ratio $R^C$ increases linearly,
hinting a repulsive force in this channel.
Moreover, after a starting increase up to $t \sim 4$,
the rectangular amplitude ratio $R^R$
demonstrates a roughly linear decrease up until $t \sim 15$,
and the signals become noisy after that,
suggesting an attractive force between pion and kaon in this channel.
These characteristics are what we expected from
the theoretical predictions~\cite{Bernard:1990kw,Sharpe:1992pp}.

\begin{figure}[htb!]
\begin{center}
\includegraphics[width=8cm,clip]{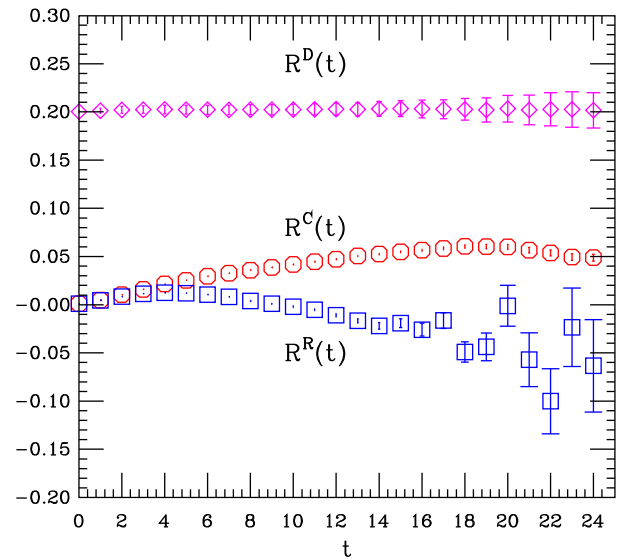}
\end{center}
\vspace{-0.3cm}
\caption{
(color online). Individual amplitude ratios $R^X(t)$ ($X=D, C$ and $R$) for
the $\pi K$ four-point correlation function measured by the moving wall
source without gauge fixing as the functions of $t$.
Direct diagram (magenta diamonds) shifted by $0.8$,
crossed diagram (red octagons) and
rectangular diagrams (blue squares).
\label{fig:ratio}
}
\end{figure}

We can observe that the crossed and rectangular amplitudes
take the same value at $t=0$, and the similar values for small $t$.
Since our analytical representations for both amplitudes are
identical at this value of $t$,
they should manifest analogously until the asymptotic $\pi K$ state is reached.
Clear signals observed up to $t = 15$ for the rectangular amplitude
demonstrate that the technique of the moving wall source without gauge fixing
used here is practical and feasible.

According to the analytical arguments in Ref.~\cite{DeGrand:1990ss},
we can readily infer that the ratio for the rectangular diagram $R^R$ has errors,
which should increase exponentially as $\displaystyle e^{m_K t}$
for large time separation.
The magnitude of the errors is in quantitatively agreement with
this theoretical prediction as displayed in Fig.~\ref{fig:error_R}.
Fitting the errors $\delta  R^R(t)$
by a single exponential fit ansatz
$\delta R^R(t) \sim  \exp(\mu_R t)$ over the range $10\le t \le 16$,
we can achieve the corresponding fitting values of $\mu_R$ with $a \mu_R = 0.358$,
which can be reasonably compared with the corresponding kaon masses $m_K$
determined in our previous work~\cite{Fu:2012zk}, which is also listed in Table~\ref{table:Disp}.
This demonstrates, on the other side,
that the technique of the moving wall source without gauge fixing
used in this work is practically feasible.
\begin{figure}[h!]
\begin{center}
\includegraphics[width=8cm,clip]{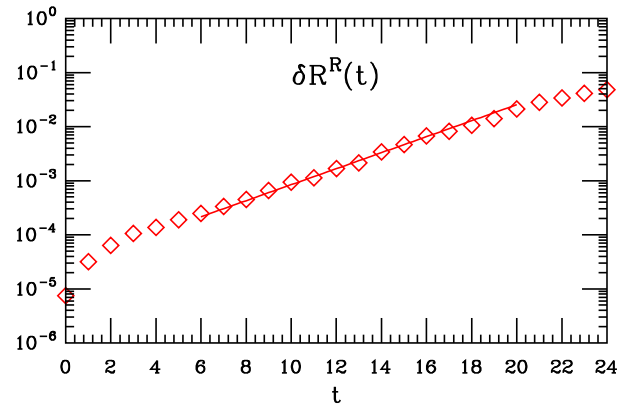}
\end{center}
\vspace{-0.3cm}
\caption{\label{fig:error_R}
(color online). The error of ratio $R^R(t)$ as a function of time separation $t$.
Solid line is a single exponential fit over the range $10\le t \le 16$.
}
\end{figure}

\subsection{Energy eigenvalues}
We calculate two eigenvalues $\lambda_n(t,t_R)$ ($n=1,2$)
for the matrix $M(t,t_R)$ denoted in Eq.~(\ref{eq:M_def})
with the reference time $t_R = 5$.
In Fig.~\ref{fig:Lambda_t}
we illustrate our lattice simulation results for $\lambda_n(t, t_R) (n = 1, 2)$
in a logarithmic scale as a function of time separation $t$
along with a correlated fit to the
asymptotic form offered in Eq.~(\ref{Eq:asy}).
From these fits the desired energies $\overline{E}_n$ $(n=1,2)$
are then obtained, and
which will be employed to derive the $p$-wave scattering phase.

\begin{figure}[h]
\begin{center}
\includegraphics[width=8.0cm]{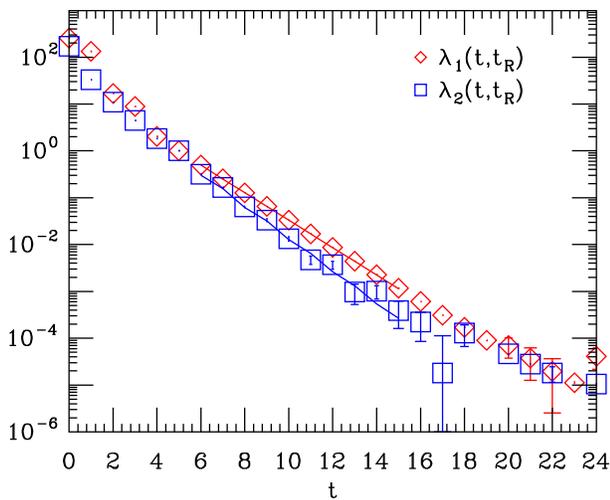}
\end{center}
\vspace{-0.3cm}
\caption{\label{fig:Lambda_t}
(color online). The eigenvalues $\lambda_1(t,t_R)$ and $\lambda_2(t, t_R)$
as a function of $t$.
Occasional points with negative central values
for the eigenvalue $\lambda_2(t, t_R)$ are not plotted.
The solid lines are the correlated fits to the asymptotic form denoted in Eq.~(\ref{Eq:asy}),
from which the energy eigenvalues $\overline{E}_n$ $(n=1,2)$ are extracted.
The lower curve ($n=2$) is slightly steeper than
the upper curve ($n=1$).
}
\end{figure}

As we noticed in Refs.~\cite{Fu:2011wc,Fu:2011xw,Fu:2012gf},
we realize that the properly extracting the energy eigenvalues
is vital to our final conclusions.
Since the PBC is imposed
on three spatial directions and the temporal direction,
we should suppress the warp-around contaminations~\cite{Fu:2011wc,Feng:2010es}.
By defining a fitting range $[t_{\rm{min}}, t_{\rm{max}}]$
and varying  the values of the minimum fitting distance $t_{\rm{min}}$
and the maximum fitting distance $t_{\rm{max}}$,
we obtain these energies in a correct manner.
In practice, we make $t_\mathrm{min}=t_R+1$ and
increase reference time $t_R$ to reduce excited contaminations~\cite{Feng:2010es}.
At the same time, we opt $t_\mathrm{max}$ to be away from the time
slice $T/2$ to reduce the warp-around effects~\cite{Feng:2010es}.
Furthermore, we extract two eigenvalues $\lambda_n (n=1,2)$
with the ``effective energy'' plots~\cite{Fu:2011wc,Feng:2010es}, 
a variant of the effective mass plots,
and they were fit to Eq.~(\ref{Eq:asy}) by changing $t_{\rm{min}}$,
and with the $t_{\rm{max}}$ either at $15$
or where the fractional statistical errors exceeded about $20\%$
for two successive time slices.
The effective energy plots as a function of $t_{\rm{min}}$
are illustrated in Fig.~\ref{fig:plateau}.
\begin{figure}[h]
\begin{center}
\includegraphics[width=8cm,clip]{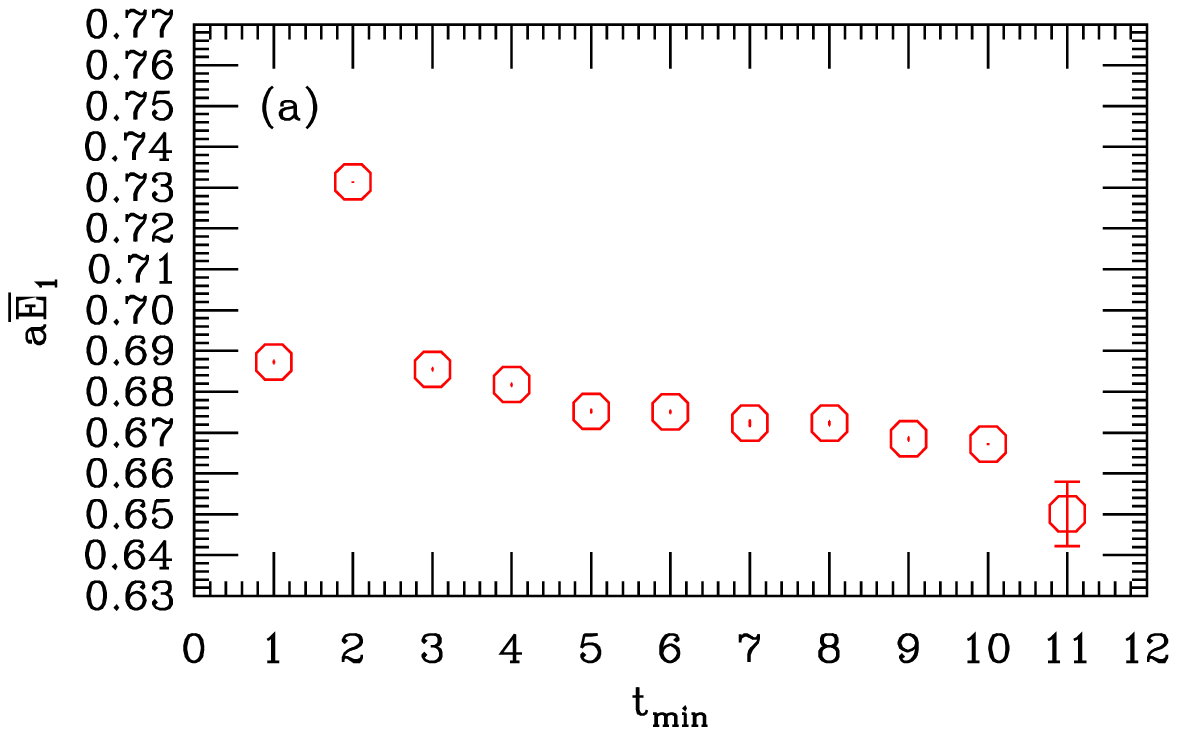}
\includegraphics[width=8cm,clip]{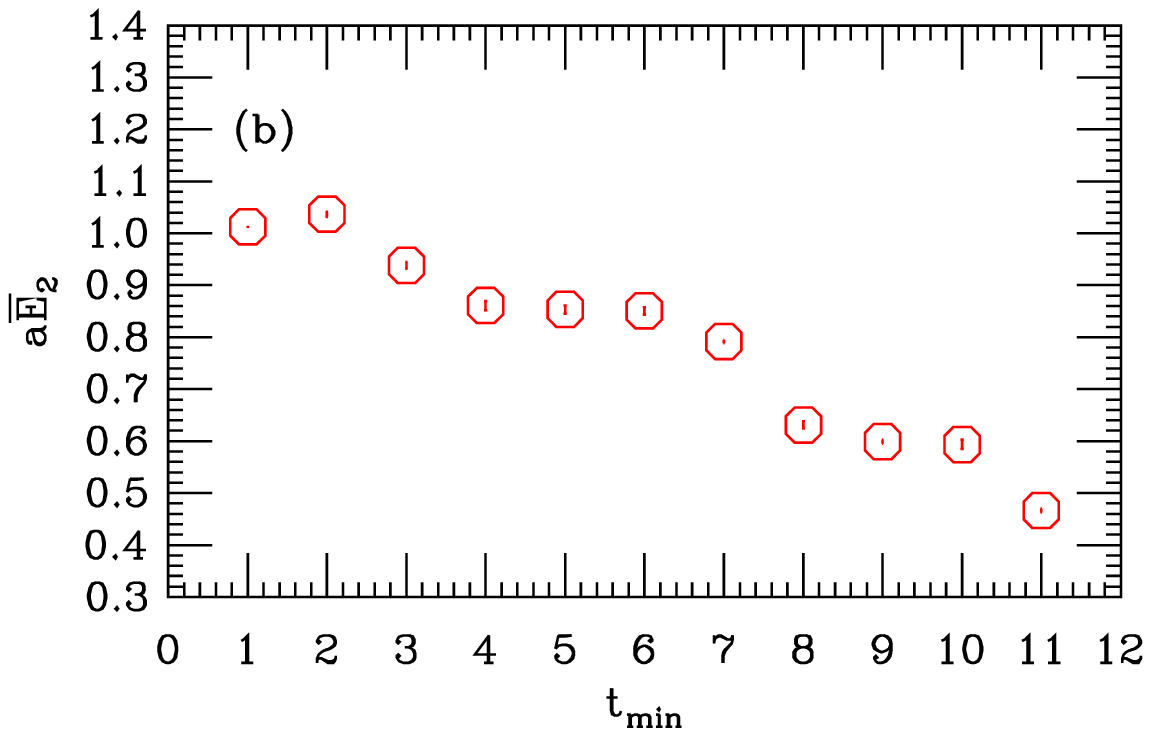}
\end{center}
\vspace{-0.5cm}
\caption{
The effective energy plots, $a \overline{E}_n$ $(n=1,2)$,
as the functions of $t_{\rm min}$.
(a) The effective energy plot for $\overline{E}_1$ and
(b) that for $\overline{E}_2$
\label{fig:plateau}
}
\end{figure}

The energy eigenvalues $\overline{E}_n (n=1,2)$ were chosen by looking for
the combination of a ``plateau'' in the effective energy plots
as the function of $t_{\rm{min}}$ and a reasonable fit quality~\cite{Fu:2011wc,Feng:2010es}.
We observed that the effective energies take only relatively small errors
within a minimum time distance region
$5 \le t_{\rm min} \le 8$ for $\overline{E}_1$ and
$5 \le t_{\rm min} \le 6$ for $\overline{E}_2$, respectively.
The fitted numbers for $\overline{E}_n$ ($n=1,2$)
along with the fitting parameters $t_R$, $t_{\mathrm{min}}$ 
and $t_{\mathrm{max}}$, fit quality $\chi^2/\mathrm{dof}$ 
are summarized in Table~\ref{tab:fitting_results}.

\begin{table}[htb]
\caption{\label{tab:fitting_results}
The fitted values of the energy eigenvalues for the ground
state ($n=1$) and the first excited state ($n=2$).
Here we tabulate the reference time $t_R$, the fitting range, $t_{\mathrm{min}}$ and $t_{\mathrm{max}}$,
the number of degrees of freedom (dof) for the fit quality $\chi^2/\mathrm{dof}$
and the fitted results for the energy eigenvalues $\overline{E}_n$ ($n=1,2$) in lattice units.
}
\begin{ruledtabular}
\begin{tabular}{cccccc}
n&$t_R$&$t_{\mathrm{min}}$&$t_{\mathrm{max}}$
& $a\overline{E}_n$ &$\chi^2/\mathrm{dof}$ \\
\hline
$1$  &  $5$  &  $6$ & $15$  & $0.67507(40)$ & $12.2/6$ \\
$2$  &  $5$  &  $6$ & $15$  & $0.8534(78)$  & $9.6/6$ \\
\end{tabular}
\end{ruledtabular}

\end{table}

The energy of the pion and kaon in non-interacting case (namely, $E_1$)
is computed from the pion mass $m_\pi$ and kaon energy $E_K$ listed in
Table~\ref{table:Disp} as $E_1 = m_\pi + E_K$.
This number is listed in Table~\ref{table:TanDel}.
We distinctly observe that $\overline{E}_1 < E_1 < \overline{E}_2$,
which mean that the $p$-wave scattering phase  for $\lambda_1(t,t_R)$
and $\lambda_2(t,t_R)$ is positive and negative, respectively.
This evidently reveals that there  exist a  resonance between $\overline{E}_1$ and
$\overline{E}_2$.
\begin{table}[h]
\caption{ \label{table:TanDel}
Summary of the energy eigenvalues $\overline{E}_n$ ($n=1,2$) and
$p$-wave scattering phase $\delta_1$ for $\pi K$ system in a cubic box.
$E_1$ is the energy of the free pion-kaon system.
$\overline{E}_n$($n=1,2$) are obtained from the correlated fits to
eigenvalues $\lambda_n(t,t_R)$ ($n=1,2$).
The invariant mass $\sqrt{s}$, the scattering momentum $k$
and the $p$-wave scattering phase $\delta_1$
derived through the expression
(\ref{eq:Disp_Two_Cont_k}) in the continuum
are regarded as {\it Cont}, and those achieved with the equation
(\ref{eq:Disp_Two_Lat_k}) on the lattice
are viewed as {\it Lat}.
The scattering momentum $k_0$ is computed by
$k_0^2 = 1/4\times(\sqrt{s} + (m_\pi^2 - m_K^2)/\sqrt{s})^2 - m_\pi^2$.
All values with the mass dimension are in lattice units.
}
\begin{ruledtabular}
\begin{tabular}{lllll}
       & \multicolumn{1}{c}{ $n=1$ } &
       & \multicolumn{1}{c}{ $n=2$ } &  \\
\hline
$E_n$            &  $ 0.67968(51) $  &
                 & \multicolumn{1}{c}{-----} &  \\
$\overline{E}_n$ &  $ 0.67507(40) $  &
                 &  $ 0.8534(78) $      \\
\hline
& \multicolumn{1}{c}{Cont} & \multicolumn{1}{c}{Lat}
& \multicolumn{1}{c}{Cont} & \multicolumn{1}{c}{Lat}   \\
$\sqrt{s}$       &  $ 0.59751(45)$   &   $ 0.60350(45) $
                 &  $ 0.7934(84)$    &   $ 0.8004(84)  $  \\
$k^2$            &  $ 0.00588(13)$   &   $ 0.00750(14) $
                 &  $ 0.0690(30)$    &   $ 0.0729(31)  $  \\
$k_0^2$  & \multicolumn{1}{c}{-----} &   $ 0.00745(13) $
         & \multicolumn{1}{c}{-----} &   $ 0.0717(30)  $  \\
$\tan\delta_1$   &  $ 0.0294(93)$    &   $ 0.0091(22)  $
                 &  $-2.01(43)$      &   $-2.48(63)   $  \\
$\sin^2\delta_1$ &  $0.00087(55)$    &   $0.000083(39) $
                 &  $0.802(68)$      &   $0.860(62)$      \\
\end{tabular}
\end{ruledtabular}
\end{table}

\subsection{Finite-size effects}
\label{SubSec: Effect of finite lattice spacing }

We should pay attention to the discretisation error (or truncation error)
inherent in the special finite size formula for $\pi K$ system
denoted in~(\ref{eq:Luscher_MF})~\cite{Fu:2011xw,Fu:2011xz}.
It stems from the Lorentz transformation
from the moving frame to the center-of-mass frame.
When applying the Lorentz symmetry in the continuum limit,
we utilize the following relations~\cite{Fu:2011xw,Fu:2011xz},
\begin{eqnarray}
\label{eq:Disp_Two_Cont_k}
\sqrt{s}  &=& \sqrt{ E_{MF}^2 - p^2 }, \cr
k^2  &=& \frac{1}{4}
\left( \sqrt{s} + \frac{m_\pi^2 - m_K^2}{\sqrt{s}} \right)^2 - m_\pi^2  ,
\end{eqnarray}
in the Lorentz transformation for the invariant mass $\sqrt{s}$,
the energy of the $\pi K$ system in the moving frame $E_{MF}$
and the scattering momentum $k$.
Nevertheless, the discretisation effects definitely
violate the Lorentz symmetry on the lattice
and equation~(\ref{eq:Disp_Two_Cont_k})
is only effective up to the truncation errors.

Following the recommendations in Ref.~\cite{Fu:2011xw},
we calculate the invariant mass $\sqrt{s}$ and the scattering momentum $k$
from the energy in the moving frame $E_{MF}$ of $\pi K$ system using
\begin{eqnarray}
\hspace{-0.5cm}\cosh( \sqrt{s} ) \hspace{-0.1cm}&=&\hspace{-0.1cm}
\cosh(E_{MF}) - 2\sin^2\left(\frac{p}{2} \right)   , \cr
\hspace{-0.5cm}2\sin^2 (k/2)      \hspace{-0.1cm}&=&\hspace{-0.1cm}
\cosh\left( \frac{\sqrt{s}}{2} + \frac{m_\pi^2 \hspace{-0.05cm}-\hspace{-0.05cm} m_K^2}{2\sqrt{s}} \right)
-\cosh(m_\pi)  ,
\label{eq:Disp_Two_Lat_k}
\end{eqnarray}
and derive the $p$-wave scattering phase $\delta_1$
by inserting the scattering momentum $k$ into the finite-size formula
in Eq.~(\ref{eq:Luscher_MF}).
We have justified these formula in Ref.~\cite{Fu:2011xw},
and we will employ them  in this work to investigate the discretisation effects.

To grasp these discretisation effects quantitatively,
in the present study we compute the invariant mass $\sqrt{s}$
and the scattering momentum $k$ not only from the energy momentum relation
in the continuum (\ref{eq:Disp_Two_Cont_k})
but also from that on the lattice (\ref{eq:Disp_Two_Lat_k}),
and subsequently  extract the $p$-wave scattering phase $\delta_1$.
People usually regard the disparity stemming from two options of
the energy momentum relations as the discretisation error.
Since it is a kind of truncation error,
it is  expected to smaller if we employ the lattice ensemble
with smaller lattice space $a$,
of course it should be disappeared in the continuum limit.
The results for the $p$-wave scattering phase $\delta_1$
along with the invariant mass $\sqrt{s}$
and the scattering momentum $k$
are summarized in Table~\ref{table:TanDel}.

\subsection{Extraction of the scattering phase and decay width}
\label{SubSec:Scattering Phase Shift and Decay Width }
The noticeable differences in the invariant mass $\sqrt{s}$
and scattering momentum $k$ because of the discretisation effects
are obviously observed from Table~\ref{table:TanDel}.
Moreover, the differences for the $p$-wave scattering phase $\delta_1$
due to the discretization effects
are impressive, and can be comparable with the statistical errors,
even considerably larger than its statistical error for the $n=1$ case.
These characteristics are visualized in Fig.~\ref{fig:sin2Del},
where the $p$-wave scattering phase $\sin^2 \delta_1$ is displayed instead~\cite{Aoki:2011yj,Fu:2011xw}.
We notice, from Table~\ref{table:TanDel},
that the numerical value of the $p$-wave scattering phase $\delta_1$
at the invariant mass $\sqrt{s}< m_{K^\ast}$ ($am_{K^\ast} = 0.7757\pm0.0070$)
is positive due to an attractive interaction,
and that at $\sqrt{s}> m_{K^\ast}$ is negative
owing to a repulsive interaction.
These features indicate a resonance at a mass around the $K^\ast$ mass.

In principle, it is a piece of cake to extract the $K^\ast(892)$ meson decay width
through fitting the $p$-wave scattering phase shift data
with the effective range formula directly~\cite{Aoki:2011yj,Fu:2011xw}.
Nevertheless, in this work we studied with the quark mass which is
larger than its nature value
and the kinematic factor in the decay width
depends clearly upon the quark mass~\cite{Nebreda:2010wv},
thus an extrapolation is indispensable~\cite{Aoki:2011yj,Fu:2011xw}.
However, we carried out a lattice calculation with one set of the quark mass
in this exploratory investigation,
therefore, we have no choice but to adopt an alternative method~\cite{Aoki:2011yj,Fu:2011xw}.
As we explained in section~\ref{SubSec:Scattering_Phase},
the resonant characteristic of the $p$-wave scattering phase $\delta_1$
are  parameterized with the coupling constant $g_{ K^\ast\pi K}$,
\begin{equation}
\tan\delta_1 = \frac{g_{ K^\ast\pi K}^2}{6\pi}\frac{k^3}{\sqrt{s}(M_R^2-s)}  ,
\label{eq:tanDel_g}
\end{equation}
where $M_R$ is the resonance mass.

According to the elaborations in Refs.~\cite{Nebreda:2010wv,Chen:2012rp},
we can fairly suppose that the coupling constant $g_{ K^\ast\pi K}$
changes quite slowly and smoothly with the quark mass.
Therefore, the equation~(\ref{eq:tanDel_g}) enables us
to solve for two unknown quantities, that is,
the coupling constant $g_{ K^\ast\pi K}$,
and the resonance mass $M_R$~\cite{Aoki:2011yj,Fu:2011xw}.

According to the discussions in Refs.~\cite{Aoki:2011yj,Fu:2011xw},
in practice, we usually employ the scattering momentum $k_0$ instead of $k$
when applying Eq.~(\ref{eq:tanDel_g}).
In Table~\ref{table:TanDel},
we provide the scattering momentum $k_0$ calculated by
$k_0^2 = 1/4 \times (\sqrt{s} + (m_\pi^2-m_K^2)/\sqrt{s})^2-m_\pi^2$
in addition to $k$.
We can observe that the difference between $k$ and $k_0$ is not significant,
and we can neglect this systemic error for the present study~\cite{Aoki:2011yj,Fu:2011xw}.

When we utilize the energy-momentum relations (\ref{eq:Disp_Two_Cont_k})
in the continuum,
the lattice simulation results of the coupling constant $g_{ K^\ast\pi K}$
and the resonance mass $M_R$ solved by Eq.~(\ref{eq:tanDel_g}) are
\begin{eqnarray}
\label{eq:FinalR_Cont}
g_{ K^\ast\pi K}  &=&  11.73 \pm 2.08,  \cr
M_R               &=&  0.739(20) ,  \cr
M_R / m_{K^\ast}  &=&  0.953(28) ,
\end{eqnarray}
where the $K^\ast$ meson mass $m_{K^\ast}$ is
quoted from our previous work~\cite{Fu:2012zk}.

On the other hand, if we adopt the energy momentum
relations (\ref{eq:Disp_Two_Lat_k}) on the lattice,
we gain the simulation results as
\begin{eqnarray}
\label{eq:FinalR_Lat}
g_{ K^\ast\pi K}   &=&  6.38(78)    ,   \cr
M_R                &=&  0.7873(97)  ,   \cr
M_R / m_{K^\ast}   &=&  1.015(16)   .
\end{eqnarray}
This obtained value of the coupling constant $g_{ K^\ast\pi K}$
in lattice case is in fair agreement with
$g_{ K^\ast\pi K} \approx 5.5$, which are obtained by Nebreda and Pel\'aez
from the residue of the amplitude at the pole position in Ref.~\cite{Nebreda:2010wv}.
Moreover, it is in reasonable agreement
with the experimental observable $g_{ K^\ast\pi K} = 5.64(35)$
evaluated from the PDG estimations of
the decay width $\Gamma_{K*} = 50.8(9)$~MeV~\cite{Beringer:1900zz}
within the statistical error.

\begin{figure}[h]
\begin{center}
\includegraphics[width=8.0cm]{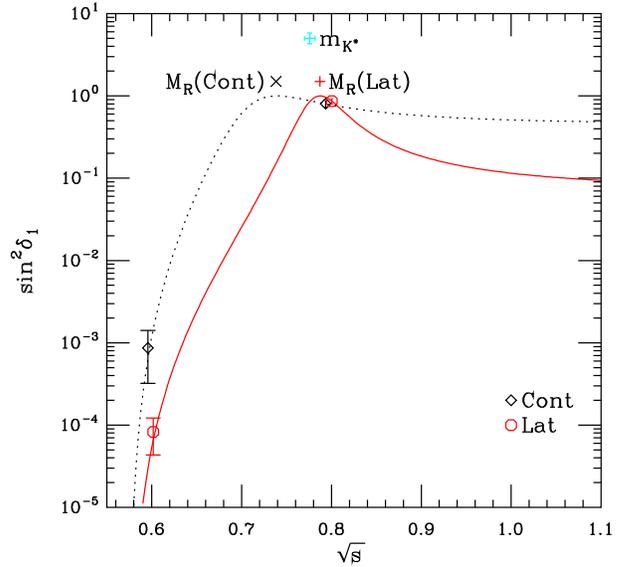}
\end{center}
\vspace{-0.4cm}
\caption{\label{fig:sin2Del}
(color online). The scattering phase $\sin^2\delta_1$ and the
positions of the $K^\ast$ meson mass $m_{K^\ast}$
and resonance mass $M_R$.
The simulation results
achieved with the energy-momentum expressions
in the continuum (\ref{eq:Disp_Two_Cont_k}) are viewed as ${\bf Cont}$
and those with the relations
on the lattice (\ref{eq:Disp_Two_Lat_k}) as {\bf Lat}.
The two lines are achieved by Eq.~(\ref{eq:tanDel_g})
with the quantities $g_{ K^\ast\pi K}$ and $M_R$
provided in Eq.~(\ref{eq:FinalR_Cont}) and Eq.~(\ref{eq:FinalR_Lat}), respectively.
The abscissa is in lattice units.
}
\end{figure}

In Fig.~\ref{fig:sin2Del}, we illustrate the curves for $\sin^2\delta_1$
achieved by equation~(\ref{eq:tanDel_g})
with the coupling constant $g_{ K^\ast\pi K}$ and
the resonance mass $M_R$ provided in Eq.~(\ref{eq:FinalR_Cont})
and Eq.~(\ref{eq:FinalR_Lat}), respectively.
The positions of the resonance mass $M_R$,
are courteously marked in Fig.~\ref{fig:sin2Del} for the two cases
(black cross and red plus for the continuum and lattice cases, respectively).
For visualized comparisons,
we  mark the $K^\ast(892)$ mass $m_{K^\ast}$ with fancy cyan plus as well.
We can observe that the resonance mass $M_R$ for lattice case
is in reasonable accordance  with
the $K^\ast(892)$ mass $m_{K^\ast}$.

Supposing that  the dependence of the coupling constant $g_{ K^\ast\pi K}$
on the quark mass is ignorable~\cite{Nebreda:2010wv,Chen:2012rp},
we can roughly estimate the $K^\ast(892)$ meson decay width
at the physical quark mass as
\begin{equation}
\Gamma^{\rm phy} =
\frac{g_{ K^\ast\pi K}^2}{6\pi}
\frac{ (k^{\rm phy})^3 }{(m_{K^\ast}^{\rm phy})^2}  ,
\end{equation}
where $m_{K^\ast}^{\rm phy}=891.66(26)$ MeV is
the physical $K^\ast(892)$ meson mass,
which we take from the current PDG~\cite{Beringer:1900zz},
and the scattering momentum $k^{\rm phy}$ at the physical point
is calculated by
$$
(k^{\rm phy})^2 = \frac{1}{4}
\left(  m_{K^\ast}^{\rm phy} + \frac{(m_\pi^{\rm phy})^2 -(m_K^{\rm phy})^2}{ m_{K^\ast}^{\rm phy}} \right)^2 - (m_\pi^{\rm phy})^2 ,
$$
where $m_\pi^{\rm phy}$ is physical pion mass
($m_\pi^{\rm phy} = 139.57018(35)$ MeV)
and $m_K^{\rm phy}$ is physical kaon mass
($m_K^{  \rm phy} = 493.677(13)$ MeV) which are quoted from the PDG~\cite{Beringer:1900zz}.
This produces
\begin{equation}
\label{eq:FinalR_Gamm_Cont}
\Gamma^{\rm phy}  = (219  \pm 39) \, {\rm MeV} \,
\end{equation}
where we utilize the simulation result given in Eq.~(\ref{eq:FinalR_Cont}).
On the other hand,
if we use these given in Eq.~(\ref{eq:FinalR_Lat}), it yields
\begin{equation}
\label{eq:FinalR_Gamm_Lat}
\Gamma^{\rm phy} = (64.9 \pm 8.0)  \, {\rm MeV} .
\end{equation}
The estimation in lattice case given in Eq.~(\ref{eq:FinalR_Gamm_Lat})
is in fair agreement with the corresponding PDG quantity
for the $ K^\ast \to \pi K$ decay width, namely,
$\Gamma_{K^\ast} = 50.8 \pm 0.9$~MeV.
We can observe that the difference stemming from
our two options of the energy-momentum relations
is much larger than the statistical error, which indicate
that the proper correction of the discretization errors
is absolutely necessary.

These results are quite stimulating,
considering that we make a big assumption
that the coupling constant  $g_{ K^\ast\pi K}$ is independent upon the quark mass,
and we carry out an extrapolation, etc.
Anyway, one thing greatly comforts us is that we use the pion mass (about $240$~MeV)
which is pretty close to its realistic value (about $140$~MeV),
so we don't carry out a long extrapolation.

\section{Conclusions and outlooks}
\label{Sec:Conclusions}
In the present work, we have carried out a lattice QCD computation
of the $p$-wave $\pi K$ scattering phase
in the $I = 1/2$ channel near the $K^\ast(892)$  resonance region
in the moving frame with total non-zero momentum 
for the MILC ``medium'' coarse ($a\approx0.15$ fm) lattice
ensemble with the $N_f=2+1$ flavors of the Asqtad
improved staggered fermions.
We employed the technique with the moving wall source
without gauge fixing~\cite{Fu:2011wc} introduced by Kuramashi et al.
in Refs.~\cite{Kuramashi:1993ka,Fukugita:1994ve}
to calculate all the three diagrams classified in Ref.~\cite{Nagata:2008wk}
with high precision.
We have exhibited that the lattice QCD computation of
the $p$-wave scattering phase for the $I=1/2$ $\pi K$ system
and then the estimation of
the decay width of $K^\ast(892)$ meson are feasible
even with our present limited computing resources.
The scattering phase data clearly reveals
a resonance at a mass around the $K^\ast(892)$ meson mass
obtained in our previous study~\cite{Fu:2012zk}.
Most of all, we extracted the $K^\ast(892)$ meson decay width
from the scattering phase data and demonstrated that
it is reasonably comparable with the $K^\ast(892)$ meson decay width
quoted from PDG within the statistical error.

We have adopted the effective range formula, which allows
us to exploit the effective $ K^\ast \to \pi K$ coupling constant
$g_{ K^\ast\pi K}$ to extrapolate
from our lattice simulation point $(m_\pi+m_K)/m_{K^\ast} = 0.7388$
to the physical point $(m_\pi+m_K)/m_{K^\ast} = 0.7102$,
assuming that the coupling constant  $g_{ K^\ast\pi K}$
is independent of the quark mass.
This is a crude estimation,
a more rigorous computation of the decay width is highly desirable.
As we pointed out above, the decay width can be reliably estimated
from the energy dependence of the scattering phase data
by fitting the BWRF
if we carry out the lattice simulations close to the physical quark mass
and obtain several energies near the resonance mass.
We will keep on enthusiastically requesting for the possible computational allocations
to fulfil this valuable work.

Nevertheless, we should bear firmly in mind that some critical issues
should be resolved in the more convincing calculation.
One is to reduce the discretization errors,
which, we illustrated in the previous section, are
significantly larger than the corresponding statistical errors.
A na\"ive way to handle this question is to
utilize a lattice ensemble closer to the continuum limit.
Another challenging and stimulating topic is to
suppress the contaminations of the $p$-wave scattering phase
from the $d$-wave scattering phase or higher,
which we preliminarily touched on this topic for the $\pi K$ system
in Ref.~\cite{Fu:2011xz},
see more valuable discussions in Ref.~\cite{Leskovec:2012gb}.
Moreover, the comprehensive investigations on
the lattice size dependence of the scattering phase
is absolutely fascinating.
Nevertheless, all of these open questions are beyond the scope of this paper
since this will demand a huge amount of computing allocations.
We postpone these expensive tasks in our future study.

This work concentrated on the $p$-wave scattering phase
only at two energies for a single lattice ensemble.
Since we had only a small number of energies at hand,
it becomes quite difficult to reliably map out the resonance region.
Therefore, our current lattice results are not
comparable with the experimentally measured quantities.
Although a reliable derivation of the $K^\ast(892)$ resonance parameters
from the lattice is absolutely big challenging and most prospective,
our rudimentary work reported here can be still viewed as
an important conceptual study, and the techniques employed
here will be helpful and useful for other resonances
such as the $D^\ast$,  possibly even for some exotic hadrons.

\section*{Acknowledgments}
We deeply appreciate the MILC Collaboration
for supplying us the Asqtad lattice ensemble and MILC codes.
We should thank NERSC (National Energy Research Scientific Center)
for providing the convenient platform to download the MILC gauge configurations
and Massimo Di Pierro for his Python toolkits.
The authors sincerely thank Carleton DeTar
for his encouraging and critical comments and supplying us the fitting software.
We especially thank Eulogio Oset for his enlightening
and constructive comments and corrections.
We are grateful to Hou Qing for his supports.
Numerical calculations for this paper were carried out at AMAX,
CENTOS and HP workstations in the Institute of
Nuclear Science and Technology, Sichuan University.

\appendix
\section{The numerical evaluation of the $\mathcal{Z}^{ \mathbf d }_{20}(1;q^2)$ function }
\label{appe:zeta}
In this appendix we follow the original derivations and notations
in Refs.~\cite{Fu:2011xw,Yamazaki:2004qb,Leskovec:2012gb}
to provide one simple approach
for the numerical evaluation of the zeta function
$\mathcal{Z}^{ \mathbf d }_{20}(s;q^2)$ defined in Eq.~(\ref{zetafunction_MF})
in the moving frame for the arbitrary value of $q^2$.

The definition of the zeta function
$\mathcal{Z}_{20}^{\mathbf d}(s;q^2)$
appeared in Eq.~(\ref{zetafunction_MF}) is
\begin{equation}
\mathcal{Z}^{ \mathbf d }_{ 20 } ( s ; q^2 ) = \sum_{ {\mathbf r} \in P_{\mathbf d} }
\frac{ {\cal Y}_{20}(\mathbf{r}) }{( r^2 - q^2 )^s }  ,
\label{eq:Z00d_appendix}
\end{equation}
where ${\cal Y}_{lm}(\mathbf{r})\equiv r^l Y_{lm}(\Omega_r)$.
$\Omega_r$ represents the solid angles $(\theta, \phi)$ of $\mathbf{r}$ in spherical coordinates and the $Y_{lm}$ are the spherical harmonic functions,
and the summation for ${\mathbf r}$ is taken over the set
\begin{equation}
\label{eq:app:pset}
P_{\mathbf d} = \left\{ {\mathbf r} \left|  {\mathbf r} = \vec{\gamma}^{-1}
\left({\mathbf n}+\frac{\alpha}{2} {\mathbf d} \right) , \quad
{\mathbf n}\in \mathbb{Z}^3 \right. \right\}  ,
\end{equation}
where
$$
\alpha = 1 + \frac{m_K^2-m_\pi^2}{E_{CM}^2} ,
$$
and the operation $\hat{ \gamma }^{ -1 }$ is defined in Eq.~(\ref{gamma_factor_sh}).

Without loss of generality, we first assume $q^2 > 0$,
and divide the summation in $\mathcal{Z}_{20}$ into two pieces as
\begin{equation}
\sum_{{\mathbf r} \in P_{\mathbf d}} \frac{ {\cal Y}_{20}(\mathbf{r})  }
{( r^2 - q^2 )^{s}} \hspace{-0.05cm}=\hspace{-0.2cm}
\sum_{r^2 < q^2}\frac{ {\cal Y}_{20}(\mathbf{r}) }{(r^2-q^2)^{s}}
\hspace{-0.05cm}+\hspace{-0.2cm}
\sum_{r^2 > q^2}\frac{ {\cal Y}_{20}(\mathbf{r}) }{(r^2-q^2)^{s}}  ,
\label{eq:app_zeta_1}
\end{equation}
where the summation over ${\mathbf r }$ is conducted
with ${\mathbf r }\in P_{ \mathbf d }$ denoted in Eq.~(\ref{eq:app:pset}).
The second term can be conveniently delivered in an integral expression,
\begin{widetext}
\begin{eqnarray}
\sum_{ r^2 > q^2 } \frac{ {\cal Y}_{20}(\mathbf{r}) }{( r^2 - q^2 )^{s}}
&=&
\frac{1}{ \Gamma(s) } \sum_{r^2 > q^2 } {\cal Y}_{20}(\mathbf{r})
\left[ \int_0^1        {\rm d}t \ t^{s-1} e^{ - t ( r^2 - q^2) } +
       \int_1^{\infty} {\rm d}t \ t^{s-1} e^{ - t ( r^2 - q^2) } \right] \cr
&=&
\frac{1}{\Gamma(s)}
\int_0^1 {\rm d}t  t^{s-1} e^{q^2 t}
\sum_{{\mathbf r} \in P_{\mathbf d}} {\cal Y}_{20}(\mathbf{r})
e^{-r^2 t}
\hspace{-0.1cm}-\hspace{-0.1cm}
\sum_{ r^2 < q^2 } \frac{ {\cal Y}_{20}(\mathbf{r}) }{( r^2 - q^2 )^{s}}
+
\sum_{{\mathbf r} \in P_{\mathbf d}} {\cal Y}_{20}(\mathbf{r})
\frac{ e^{ -( r^2 - q^2 ) } }{ (r^2 - q^2)^s }.
\label{eq:app_zeta_2}
\end{eqnarray}
\end{widetext}
The second term nicely counteract the first term in Eq.~(\ref{eq:app_zeta_1}).
Using the Poisson resummation formula, the first term results in
\begin{eqnarray}
\label{first_term}
\mathrm{first\ term} &=& \frac{1}{\Gamma(s)}
\int _0^1 dt \, t^{s-1} e^{tq^2}\sum_{\mathbf{n}\in \mathbb{Z}^3} f_{\mathbf{n}} ,
\cr
f_{\mathbf{n}} &\equiv& \int d^3{\mathbf x} \, {\cal Y}_{20}(\mathbf{r})
e^{-t|\mathbf{r}|^2 + i2\pi \mathbf{n}\cdot \mathbf{x}} ,
\end{eqnarray}
where
$\mathbf{r}=\hat \gamma^{-1}(\mathbf{x}+\frac{1}{2} \alpha \mathbf{d})$.
After transforming the integration variable  from $\mathbf{x}$ to $\mathbf{r}$,
and considering the relations: $d^3{\mathbf x} = \gamma d^3 \mathbf {r}$ and
$\mathbf{x} = \hat \gamma \mathbf{r} - \frac{1}{2} \alpha\mathbf{d}$,
then we can separate terms which are dependent only upon ${\mathbf r}$
$$
f_{\mathbf{n}} \equiv \gamma \, e^{-i\pi \alpha \mathbf{n\cdot d}}\
\int d^3{\mathbf r} \, {\cal Y}_{20}(\mathbf{r}) e^{-t|\mathbf{r}|^2 + i2\pi \hat{\gamma}
\mathbf{n}\cdot \mathbf{r}} .
$$
Let $\mathbf{k}\equiv \pi \hat{\gamma} \mathbf{n}$, we rewrite above equation as
$$
f_{\mathbf{n}} \equiv \gamma \, e^{-i\pi \alpha \mathbf{n\cdot d}}\
e^{ - k^2/t^2 } \,
\int d^3{\mathbf r} \, {\cal Y}_{20}(\mathbf{r}) e^{-t( \mathbf{r} - i\mathbf{k}/t)^2 } ,
$$
where $\mathbf{r}=(x,y,z)$.
Let us conduct a variable substitution, namely,
 $\mathbf{r} - i\mathbf{k}/t \to \mathbf{r}$,
we can strictly verify
\begin{eqnarray}
\int d^3{\mathbf r} \,  x^2 e^{-t( \mathbf{r} - i\mathbf{k}/t)^2 } &=&
\frac{2\pi}{t} \int_0^\infty dx  \left(x^2 - \frac{k_x^2}{t^2} \right) e^{-tx^2 } \cr
&=&
\left(\frac{\pi}{t}\right)^{3/2} \left(\frac{1}{2t}-\frac{k_x^2}{t^2}\right), \cr
\int d^3{\mathbf r} \,  y^2 e^{-t( \mathbf{r} - i\mathbf{k}/t)^2 } &=&
\left(\frac{\pi}{t}\right)^{3/2} \left(\frac{1}{2t}-\frac{k_y^2}{t^2}\right), \cr
\int d^3{\mathbf r} \, z^2 e^{-t( \mathbf{r} - i\mathbf{k}/t)^2 } &=&
\left(\frac{\pi}{t}\right)^{3/2} \left(\frac{1}{2t}-\frac{k_z^2}{t^2}\right) .
\end{eqnarray}
We finally obtain
$$
f_{\mathbf{n}} \equiv -\gamma \, e^{-i\pi \alpha \mathbf{n\cdot d}}\
e^{ - k^2/t^2 } \, \frac{\pi^{3/2}}{t^{7/2}}  {\cal Y}_{20}(\mathbf{k}) ,
$$
where ${\cal Y}_{20}(\mathbf{k}) \equiv k^2~Y_{20}(\Omega_k)$.
Now we can rewrite the first term in Eq.~(\ref{eq:app_zeta_2}) as
\begin{eqnarray}
\mathrm{first\ term} &=&
\frac{\gamma}{ \Gamma(s) }
\int_0^1  {\rm d}t \, t^{s-1} e^{t q^2}
\frac{ \pi^{\frac{3}{2}} }{t^{\frac{7}{2}}}
\sum_{ {\mathbf n } \in \mathbb{Z}^3 }
(\pi \hat{\gamma}{\mathbf n} )^2 Y_{20}(\Omega_k) \cr
&& \times e^{i \pi \alpha \, {\mathbf n} \cdot {\mathbf d} }
e^{ -  ( i\pi \hat{\gamma}{\mathbf n} )^2/t }  .
\label{eq:app_zeta_3}
\end{eqnarray}
After collecting all terms, it leads to the representation of
the zeta function  at $s=1$,
\begin{eqnarray}
\mathcal{Z}_{20}^{\mathbf d} (1; q^2) &=&
\sum_{ {\mathbf r} \in P_{\mathbf d} } r^2 Y_{20}(\Omega_r)
\frac{ e^{ -(r^2-q^2) } }{r^2-q^2}  \cr
&& -
\int_0^1  {\rm d}t \, e^{t q^2}
\frac{ \pi^{\frac{3}{2}} }{t^{\frac{7}{2}}}
\sum_{ {\mathbf n } \in \mathbb{Z}^3 }
(\pi \hat{\gamma}{\mathbf n} )^2 Y_{20}(\Omega_k) \cr
&& \times e^{-i \pi \alpha   {\mathbf n} \cdot {\mathbf d} }
e^{ -(\pi \hat{\gamma}{\mathbf n} )^2/t } .
\label{eq:app_zeta_s=1}
\end{eqnarray}

Now let us consider the case of $q^2 \le 0$.
As a matter of fact, there is no need for us to divide the summation
in $\mathcal{Z}_{20}^{\mathbf d}(s; q^2)$ into two parts,
and can be  delivered in an integral expression as well~\cite{Fu:2011wc,Fu:2011xw}.
Conducting the same procedures, we obtain the same expression
in Eq.~(\ref{eq:app_zeta_s=1}).
Therefore, equation~(\ref{eq:app_zeta_s=1}) is applicable for the arbitrary value of $q^2$.

Plugging in ${\mathbf d} = (0,0,1)$ into equation~(\ref{eq:app_zeta_s=1}),
we arrive at the representation of the zeta function
$\mathcal{Z}_{20}^{\mathbf d}(s; q^2)$ in Eq.~(\ref{zetafunction_MF})
devoted for the current work
\begin{eqnarray}
\mathcal{Z}_{ 20 }^{\mathbf d} (1; q^2) &=&
\sum_{ {\mathbf r} \in P_{\mathbf d} } r^2 Y_{20}(\Omega_r)
\frac{ e^{ -(r^2-q^2) } }{r^2-q^2} \cr
&& - \int_0^1  {\rm d}t  \, e^{t q^2}
\frac{ \pi^{\frac{3}{2}} }{t^{\frac{7}{2}}}
 \sum_{ {\mathbf n } \in \mathbb{Z}^3 }
(\pi \hat{\gamma}{\mathbf n} )^2 Y_{20}(\Omega_k) \cr
&& \times \cos( \pi \alpha   {\mathbf n} \cdot {\mathbf d} )
e^{ - ( \pi \hat{\gamma}{\mathbf n} )^2/t } ,
\end{eqnarray}
where only the real part of the zeta function
$\mathcal{Z}_{ 20 }^{\mathbf d} (1; q^2)$ is survived.

I also note that the general numerical evaluation
of the zeta function $\mathcal{Z}_{lm}^{\mathbf d}(s; q^2 )$
has been derived in Refs.~\cite{Fu:2011xz,Leskovec:2012gb}.
We numerically compared both these representations of
the zeta  function $\mathcal{Z}_{20}(1;q^2)$ with this representation,
and found that they are reasonable consistent.


%
\end{document}